\documentclass[aps,pre,reprint,superscriptaddress,showpacs,preprintnumbers,amsmath,amssymb,nofootinbib]{revtex4-2}
\usepackage[dvipdfmx]{graphicx}
\usepackage{dcolumn}
\usepackage{bm}
\usepackage{amssymb, amsmath, amsfonts}
\usepackage{lipsum}
\usepackage{color}
\usepackage{multirow}
\usepackage{appendix}
\DeclareGraphicsExtensions{{.pdf}}
\begin{document}
\newcommand{\noter}[1]{{\color{red}{#1}}}
\newcommand{\noteb}[1]{{\color{blue}{#1}}}
\newcommand{\phic}{{V_{ij}(r_{ij}^{\rm c})}}
\newcommand{\dphic}{{V^{\prime}_{ij}(r_{ij}^{\rm c})}}
\newcommand{\ddphic}{{V^{\prime\prime}_{ij}(r_{ij}^{\rm c})}}

\widetext

\title{Potential energy landscape picture of zero-temperature avalanche criticality governing dynamics in supercooled liquids
}

\author{Norihiro Oyama}
\affiliation{Toyota Central R\&D Labs., Inc., {Nagakute 480-1192}, Japan}

\author{Yusuke Hara}
\email{Yusuke.Hara.ys@mosk.tytlabs.co.jp}
\affiliation{Toyota Central R\&D Labs., Inc., {Nagakute 480-1192}, Japan}

\author{Takeshi Kawasaki}
\affiliation{D3 Center, The University of Osaka, Toyonaka, Osaka 560-0043, Japan}
\affiliation{Department of Physics, The University of Osaka, Toyonaka, Osaka 560-0043, Japan}
\affiliation{Department of Physics, Nagoya University, Nagoya 464-8602, Japan}

\author{Kang Kim}
\affiliation{Division of Chemical Engineering, Graduate School of Engineering Science, The University of Osaka, Toyonaka, Osaka 560-8531, Japan}

\date{\today}
\begin{abstract}
Supercooled liquids are metastable states realized by suppressing crystallization below the melting temperature. 
While it is well established that their dynamics slow down dramatically and become spatially heterogeneous upon cooling, the microscopic origin of these nontrivial glassy phenomena remains a matter of active debate.
In the present study, by means of molecular dynamics simulations, we first demonstrate that nontrivial slow dynamics, such as structural relaxation and dynamical heterogeneity, can be consistently described within a zero-temperature avalanche criticality picture.
Since this finding suggests that the potential energy landscape plays a crucial role in determining the dynamics, we further quantify the potential energy landscape from three distinct perspectives. 
Based on these analyses, we propose a potential-energy-landscape picture of avalanche criticality that is consistent with various previous studies. 
Our proposed picture explains in a unified manner previously unexplained observations near the mode-coupling transition, such as the saturation of the dynamical susceptibility and the localization of unstable modes in saddle configurations.

\end{abstract}

\maketitle

\section{Introduction}
For many soft matter systems, ranging from simple molecular liquids \cite{whitakerKineticStabilityHeat2015} to colloidal dispersions \cite{weeksThreeDimensionalDirectImaging2000,brambillaProbingEquilibriumDynamics2009,mattssonSoftColloidsMake2009}, polymers \cite{keddieInterfaceSurfaceEffects1994,chatInfluenceTacticityGlassTransition2021}, alloys \cite{pekerHighlyProcessableMetallic1993,shengAtomicPackingShorttomediumrange2006}, and cytoplasm \cite{parryBacterialCytoplasmHas2014,nishizawaUniversalGlassformingBehavior2017,oyamaGlassyDynamicsModel2019}, rapid quenching suppresses crystallization and leads to a supercooled liquid state even below the melting point \cite{cavagnaSupercooledLiquidsPedestrians2009}.
In supercooled liquids, particle mobility becomes spatially heterogeneous, leading to the emergence of mesoscale domains consisting of highly mobile and nearly immobile regions, whose size grows as the temperature is lowered~\cite{yamamotoKineticHeterogeneitiesHighly1997,yamamotoHeterogeneousDiffusionHighly1998,yamamotoDynamicsHighlySupercooled1998}.
This phenomenon, known as dynamical heterogeneity (DH), is known to be ubiquitously observed across a wide range of systems~\cite{trachtLengthScaleDynamic1998,freyLiquidlikeStressdrivenDynamics2025,kegelDirectObservationDynamical2000,weeksThreeDimensionalDirectImaging2000,berthierDynamicalHeterogeneitiesGlasses2011,angeliniGlasslikeDynamicsCollective2011,kanayamaRelationDynamicHeterogeneities2022}, and has therefore been extensively studied to date. 
Nevertheless, even for simplified model systems, the microscopic origin of DH remains elusive, making it one of the central open problems in the field.

Numerous theoretical and numerical attempts have been made to elucidate the origin of DH. 
One representative example is the random first-order transition (RFOT) theory \cite{bouchaudAdamGibbsKirkpatrickThirumalaiWolynesScenarioViscosity2004,kirkpatrickScalingConceptsDynamics1989}, which seeks to describe the dynamics of finite-dimensional systems based on predictions from replica theory, a thermodynamic mean-field theory
\footnote{
In mean-field models such as the p-spin model \cite{gardnerSpinGlassesPspin1985,crisantiSphericalpspinInteractionSpin1992,kirkpatrickSpininteractionSpinglassModels1987} and the Potts model \cite{grossMeanfieldTheoryPotts1985}, a random first-order transition associated with one-step replica symmetry breaking is observed. 
Building on these mean-field predictions, the RFOT theory was proposed as a theoretical framework to describe the various nontrivial dynamical properties exhibited by supercooled liquids in finite-dimensional systems~\cite{biroliRFOTTheoryGlasses2024}.
}.
In particular, the RFOT theory introduces a static length scale by considering the free-energy balance associated with cooperatively rearranging regions (CRRs), with the CRR size defining the characteristic length.
This length scale is predicted to increase upon cooling~\cite{kirkpatrickScalingConceptsDynamics1989}, which naturally leads to the expectation of a correlation with the characteristic length of DH.
In numerical simulations, the correlation length associated with the size of a CRR is quantified by the point-to-set length\cite{bouchaudAdamGibbsKirkpatrickThirumalaiWolynesScenarioViscosity2004}.
As another theoretical framework based on static structure, the frustration-limited domain theory (FLDT) has been proposed~\cite{kivelsonViewpointModelTheory2013}.
In this theory, the growth of order associated with locally favored structures is considered to be the key mechanism underlying dynamical slowing down.
Due to geometric frustration associated with locally favored structures, domain growth remains finite and criticality is avoided.
In the FLDT, the domain size of locally favored structures is expected to increase upon cooling and to underlie DH.
Particle-based simulations have reported that both the point-to-set length associated with RFOT \cite{hockyGrowingPointtoSetLength2012,berthierStaticPointtosetCorrelations2012,berthierEfficientMeasurementPointtoset2016} and the characteristic length scale of locally favored structures, which serves as the starting point of the FLDT framework \cite{coslovichUnderstandingFragilitySupercooled2007,coslovichLocallyPreferredStructures2011}, are shorter than the length scale of DH of systems without clear medium-range crystalline orders \cite{tahGlassTransitionSupercooled2018a}.
These findings suggest that these theories may be insufficient to fully account for the origin of DH in such systems.
We also mention that, in recent years, the central role of growing static length scales for explaining the glass transition has been questioned, leading to active debate~\cite{wyartDoesGrowingStatic2017a,berthierCanGlassTransition2019,biroliRFOTTheoryGlasses2024}.
As counterevidence, recent works have proposed an alternative scenario in which the dynamical slowing down is primarily governed by the growth of local activation energies, as suggested by elastic models \cite{picaciamarraLocalVsCooperative2024,jiRoleExcitationsSupercooled2025}.

Similarly, a line of studies has searched for static structural orders that strongly correlate with the spatial structure of DH~\cite{tanakaCriticallikeBehaviourGlassforming2010,kawasakiCorrelationDynamicHeterogeneity2007,kawasakiStructuralOriginDynamic2010}.
In particular, the order parameters proposed by Tong and Tanaka~\cite{tongRevealingHiddenStructural2018,tongStructuralOrderGenuine2019} exhibit strong correlations with DH across various systems when appropriately coarse-grained over suitable length scales.
The correlation tends to become higher in lower temperatures, and it reaches more than 0.9 in terms of the Spearman's correlation coefficient.
Moreover, inspired by the success of these heuristically designed structural order parameters, there has been growing interest in extracting structural order parameters using machine learning.
Early studies investigated whether neural networks could learn the correspondence between structure and dynamics and thereby predict dynamical behavior solely from structural information \cite{bapstUnveilingPredictivePower2020,shibaBOTANBOndTArgeting2023}.
In these studies, both structural and corresponding dynamical information were fed during the training process.
More recently, it has been shown that correlations between structure and DH can be detected even by unsupervised learning methods that do not incorporate dynamical information during training \cite{swansonDeepLearningAutomated2020,oyamaWhatDeepNeural2023,liuClassificationSolidLiquid2024}.
However, in methods based on local structural order parameters \cite{tongRevealingHiddenStructural2018,tongStructuralOrderGenuine2019,liuClassificationSolidLiquid2024}, for both heuristic and data-driven approaches, it is still necessary to refer to dynamical information when optimizing the coarse-graining length scale.
Moreover, for a representative glass-forming system, the two-dimensional modified Kob-Andersen model, the Pearson correlation coefficient between the spatial distribution of DH (quantified by the so-called dynamical propensity~\cite{widmer-cooperHowReproducibleAre2004,berthierStructureDynamicsGlass2007}) and the coarse-grained structural order remains at around 0.5 even in the deeply supercooled regime near the MCT point \cite{liuClassificationSolidLiquid2024}.
These observations suggest that, at least for some systems, the dynamics may not be uniquely determined by structural information alone.

In contrast to the static-structure-based approaches reviewed so far, there also exist theoretical frameworks that place primary emphasis on dynamics.
Although such theories include other important examples, such as mode-coupling theory (MCT)~\cite{biroliDivergingLengthScale2004,biroliInhomogeneousModeCouplingTheory2006,janssenModeCouplingTheoryGlass2018} and the distinguishable-particle lattice model~\cite{zhangEmergentFacilitationBehavior2017,lulliKovacsEffectGlass2021,leeFragileGlassesAssociated2020,lulliSpatialHeterogeneitiesStructural2020,leeLargeHeatcapacityJump2021}, in this study we focus on a recent stream of research based on the dynamical facilitation picture~\cite{chackoElastoplasticityMediatesDynamical2021a,scallietThirtyMillisecondsLife2022}.
Kinetically constrained models, including the Fredrickson-Andersen model \cite{fredricksonKineticIsingModel1984}, the East model \cite{jckleHierarchicallyConstrainedKinetic1991}, and the Kob-Andersen lattice gas model \cite{kobKineticLatticegasModel1993}, are thermodynamically trivial yet display glassy dynamical slowing down and DH driven purely by kinetic constraints.
In these pioneering models, the focus was on demonstrating that nontrivial dynamical behavior can emerge solely from dynamical rules, and the adopted rules do not fully capture the physical constraints present in real systems.
Recently, motivated by insights obtained from particle-based simulations, a thermal elastoplastic model (T-EPM) incorporating more realistic rules has been proposed \cite{ozawaElasticityFacilitationDynamic2023a}.
Furthermore, using the T-EPM, Ref.~\cite{tahaeiScalingDescriptionDynamical2023a} proposed that DH can be explained in terms of avalanche criticality.
As its name suggests, the T-EPM is an extension of elastoplastic models (EPMs) to describe thermally activated relaxation processes.
Although EPMs were originally developed to describe plastic deformation in amorphous solids under shear, two key findings from particle-based simulations motivated the extension of EPMs to thermal relaxation.
The first is the observation that elastic fields, which are a main ingredient of avalanche criticality~\cite{fisherCollectiveTransportRandom1998}, are also relevant to the relaxation dynamics of supercooled liquids \cite{wuAnisotropicStressCorrelations2015,maierEmergenceLongRangedStress2017,lemaitreStructuralRelaxationScaleFree2014,chackoElastoplasticityMediatesDynamical2021a}.
The second is the observation that shear transformation zones (STZs), which are believed to be the elementary processes underlying plastic deformation under shear \cite{maloneyUniversalBreakdownElasticity2004,oyamaShearinducedCriticalityGlasses2023a}, are also present under thermal relaxation \cite{lerbingerRelevanceShearTransformations2022}.
The interaction between STZs mediated by elastic fields is precisely the picture on which conventional EPMs are based~\cite{nicolasDeformationFlowAmorphous2018}.
Plastic events described by EPMs under shear have been reported in many studies to exhibit avalanche criticality \cite{oyamaUnifiedViewAvalanche2021,oyamaInstantaneousNormalModes2021b,oyamaScaleSeparationShearInduced2024a,oyamaShearinducedCriticalityGlasses2023a,linScalingDescriptionYielding2014a,ferreroElasticInterfacesDisordered2019,liuDrivingRateDependence2016}.
Ref.~\cite{tahaeiScalingDescriptionDynamical2023a} suggested that, by introducing the concept of thermal avalanches, plastic deformation under shear and thermal relaxation may be understood within a unified framework of avalanche criticality.
This phenomenological picture assumes that DH are controlled by avalanche criticality, with the critical point located at $T=0$.
Although particle-based numerical simulations have reported finite-size effects in dynamical susceptibilities (a representative quantitative measure of DH: defined by Eq.~\ref{eq:chi4} in Sec.~\ref{sec:results1}) that may be consistent with this picture~\cite{karmakarGrowingLengthTime2009a}, their underlying physical mechanism remains elusive, and no analysis has yet been carried out from the viewpoint of avalanche criticality.

In the accompanying Letter~\cite{DHLetter}, we demonstrate, using molecular dynamics simulations, that the temperature and system-size dependence of the dynamical susceptibility in the Kob-Andersen model (KAM)~\cite{kobTestingModecouplingTheory1995e}, a canonical model of supercooled liquids, can be explained within the zero-temperature avalanche criticality picture proposed in Ref.~\cite{tahaeiScalingDescriptionDynamical2023a}.
In particular, we validated the avalanche criticality picture by demonstrating a successful finite-size scaling collapse of the dynamical susceptibility using independently determined critical exponents.
In this full paper, we present results of more comprehensive investigations from multiple perspectives aimed at deepening our understanding of the avalanche criticality observed in the slow dynamics of the supercooled-liquid state of the KAM.
We first describe the simulation setup in detail in Sec.~\ref{sec:sim}. 
We then recapitulate, in Sec.~\ref{sec:results1}, the results presented in the accompanying Letter. 
In addition, we demonstrate that the temperature and system-size dependence of the average structural relaxation dynamics can also be consistently interpreted within the same avalanche criticality picture.
Since the validity of the zero-temperature avalanche criticality picture suggests that the potential energy landscape (PEL) plays an important role in determining the dynamics, in Sec.~\ref{sec:results2} we quantitatively characterize the PEL from three different perspectives: the vibrational density of states of inherent structures, the localization properties of unstable modes associated with saddle configurations, and the potential energy of inherent structures.
In Sec.~\ref{sec:dis}, we discuss the relationship between our findings and various previous studies. 
Importantly, we propose a PEL-based interpretation of avalanche criticality that is broadly consistent with earlier works.
The proposed PEL picture also provides a natural interpretation of two nontrivial behaviors observed around the MCT transition point, which have been established but remained unexplained. 
The first is the saturation of the dynamical susceptibility~\cite{coslovichDynamicThermodynamicCrossover2018a,dasCrossoverDynamicsKobAndersen2022a}, which indicates the breakdown of avalanche criticality around $T_{\rm MCT}$.
That is, the criticality we numerically confirmed becomes irrelevant in the deeply supercooled regime and thus does not immediately suggest that the ideal glass transition takes place at exactly zero temperature.
Our PEL-based interpretation also explains this vanishing of criticality.
The second is the localization of unstable modes at saddle-point configurations~\cite{coslovichLocalizationTransitionUnderlies2019b}.
Our picture suggests that these two behaviors arise from the same underlying mechanism, indicating that their simultaneous observation near the MCT point is not a coincidence.
We further discuss the possibility that DH may not be universally accounted for by the avalanche criticality picture in other supercooled liquid models.

\section{Simulation setups}\label{sec:sim}
We perform molecular dynamics simulations of the Kob-Andersen model (KAM), a prototypical model of supercooled liquids \cite{kobTestingModecouplingTheory1995e}.
The KAM is a binary Lennard-Jones liquid inspired by the ${\rm Ni}_{80}{\rm P}_{20}$ alloy system.
The interparticle potential is given by
\begin{align}
V_{ij}=4\epsilon_{ij}\left[\left(\frac{\sigma_{ij}}{r_{ij}}\right)^{12}-\left(\frac{\sigma_{ij}}{r_{ij}}\right)^{6}\right],\label{eq:LJ}
\end{align}
where $r_{ij}$ denotes the center-to-center distance between particles $i$ and $j$. The parameters $\epsilon_{ij}$ and $\sigma_{ij}$ set the energy scale and the interaction length, respectively. 
In the KAM, these parameters are chosen in a nonadditive manner as $\sigma_{AA}=1$, $\sigma_{BB}=0.88$, $\sigma_{AB}=0.8$, $\epsilon_{AA}=1$, $\epsilon_{BB}=0.5$, and $\epsilon_{AB}=1.5$. 
Here, the subscripts $A$ and $B$ distinguish the particle species.
The cutoff distance is set to $r_{ij}^{\rm c}=2.5\sigma_{ij}$, and the number density is fixed at $\rho=N/V=1.2$, where $V$ denotes the system volume and $N$ is the total number of particles. 
The numbers of particles of each species, $N_A$ and $N_B$, are chosen as $N_A:N_B=80:20$, corresponding to the ${\rm Ni}_{80}{\rm P}_{20}$ composition.
The particle mass is taken to be $m$ for both species. 
Throughout this paper, all physical quantities are nondimensionalized by the energy unit $\epsilon_{AA}$, the length unit $\sigma_{AA}$, and the mass unit $m$. 
We consider a three-dimensional system ($d=3$).

In this study, as discussed in Sec.~\ref{sec:deloc}, we analyze particle configurations located at saddle points of the PEL. 
To obtain such saddle-point configurations, the interparticle potential is smoothed such that it continuously goes to zero up to the second-order derivative at the cutoff distance $r_{ij}^{\rm c}$.
We employ a cubic polynomial smoothing scheme~\cite{grigeraGeometricApproachDynamic2002b,coslovichLocalizationTransitionUnderlies2019b}:
\begin{align}
V_{ij}^{\rm cubic}
&= V_{ij} + B_{ij}(a_{ij}-r_{ij})^3 + C_{ij}.
\label{eq:cubic}
\end{align}
The explicit expressions for the coefficients are given by
\begin{align}
a_{ij}&=r_{ij}^{\rm c}-2\cfrac{\dphic}{\ddphic},\\
B_{ij}&=\cfrac{(\ddphic)^2}{12\dphic},\\
C_{ij}&=\cfrac{(\ddphic)^3}{216B_{ij}^2}-\phic,
\end{align}
where we define $V_{ij}^{\prime}(r_{ij})\equiv\frac{\partial V_{ij}}{\partial r_{ij}}$ and $V_{ij}^{\prime\prime}(r_{ij})\equiv\frac{\partial^2 V_{ij}}{\partial r_{ij}^2}$.

In this study, we perform numerical simulations in the canonical ensemble. It is known that, when measuring the dynamical susceptibility, which is a representative quantitative measure of DH and is defined by Eq.~\ref{eq:chi4} in Sec.~\ref{sec:results1}, qualitatively different results can be obtained depending on the choice of thermostat \cite{berthierSpontaneousInducedDynamic2007a}. 
We employ the Nosé-Hoover thermostat, which is known to allow the measurement of the total dynamical susceptibility and has been widely employed in recent studies.

In the KAM, the density is fixed at $\rho=1.2$, and therefore, in the canonical ensemble, the only free parameters are the system size (number of particles) $N$ and the temperature $T$. 
In this study, we perform simulations by varying these parameters in the ranges $200 \le N \le 1500$ and $0.41 \le T \le 1.0$.
In this study, for each system size, we first generated completely random particle configurations as initial conditions.
Simulations were started at $T_{\rm onset} \approx 1.0$, where slow dynamics characteristic of supercooled liquids begin to emerge \cite{sastrySignaturesDistinctDynamical1998b}.
The temperature was then progressively lowered, and at each temperature the final configuration was used as the initial condition for the next simulation at a slightly lower temperature.
At each temperature, we first performed equilibration runs of duration longer than twenty times the $\alpha$-relaxation time.
This was followed by production runs of the same duration for measuring statistical observables.
To obtain reliable statistical averages, we performed simulations with at least 256 independent realizations for each combination of system size $N$ and temperature $T$.
The time step used in the molecular dynamics simulations was fixed at $\Delta t = 0.005$.
The relaxation time parameter of the Nos\'e-Hoover thermostat was set to
$\tau_T = 50\,\Delta t$~\cite{coslovichDynamicThermodynamicCrossover2018a}.

It has been pointed out that, in the KAM, crystallization may become non-negligible at low temperatures even on time scales accessible to simulations
\cite{coslovichDynamicThermodynamicCrossover2018a,ingebrigtsenCrystallizationInstabilityGlassForming2019a,dasCrossoverDynamicsKobAndersen2022a,ortliebProbingExcitationsCooperatively2023}.
However, within the ranges of temperature $T$, system size $N$, and simulation duration considered in this study, excluding crystallized samples did not lead to any qualitative differences in the results.
Accordingly, we included all samples in the analyses presented in the main text.
Details of our examination of effects of crystallization are summarized in Appendix~\ref{ap:crystal}.

\section{Results: relaxational dynamics and dynamical  heterogeneity}\label{sec:results1}
In this study, we use the overlap function
\begin{align}
    Q(t) = \langle q(t) \rangle
\end{align}
as a quantitative measure of structural relaxation dynamics.
Here, $q(t)$ is defined as 
\begin{align}
    q(t) = \frac{1}{N_A} \sum_{i=1}^{N_A}
\Theta\!\left(a - \left|\boldsymbol{r}_i(t_0+t)-\boldsymbol{r}_i(t_0)\right|\right),
\end{align}
where the sum runs over only the majority $A$ particles.
$N_A \equiv 0.8N$ denotes the number of $A$ particles.
$\Theta(x)$ is the Heaviside step function.
Thus, $q(t)$ corresponds to the fraction of $A$ particles whose displacement from the reference time $t_0$ remains smaller than $a$ after a time interval $t$.
We choose $a = 0.3$, which approximately corresponds to the plateau height of the mean-squared displacement~\cite{karmakarGrowingLengthTime2009a,dasCrossoverDynamicsKobAndersen2022a}.
The angular brackets $\langle \cdots \rangle$ denote an average over independent samples as well as over the choice of the reference time $t_0$.

As a quantitative measure of DH, we introduce the following dynamical susceptibility defined from the fluctuations of $q(t)$:
\begin{align}
\chi_4(t)=N\left[\langle q^2(t)\rangle-\langle q(t)\rangle^2\right].\label{eq:chi4}
\end{align}
The dynamical susceptibility $\chi_4(t)$ is often interpreted as reflecting the size of dynamically correlated regions during the structural relaxation
\cite{toninelliDynamicalSusceptibilityGlass2005,berthierDirectExperimentalEvidence2005}.

In this section, based on these observables, we explain how the dynamics of supercooled liquids can be understood within the avalanche criticality picture. 
We first recap, in Sec.~\ref{sec:chi4}, our findings presented in the accompanying Letter~\cite{DHLetter}, where we showed that the temperature $T$ and system-size $N$ dependence of $\chi_4(t)$ can be interpreted from the avalanche criticality picture.
Next, in Sec.~\ref{sec:Q}, we show that finite-size effects also appear in $Q(t)$, and that these effects, consistent with those observed in $\chi_4$, can likewise be understood within the avalanche criticality picture.

\begin{figure*}[!ht]
\begin{center}
\includegraphics[width=.8\linewidth,angle=0]{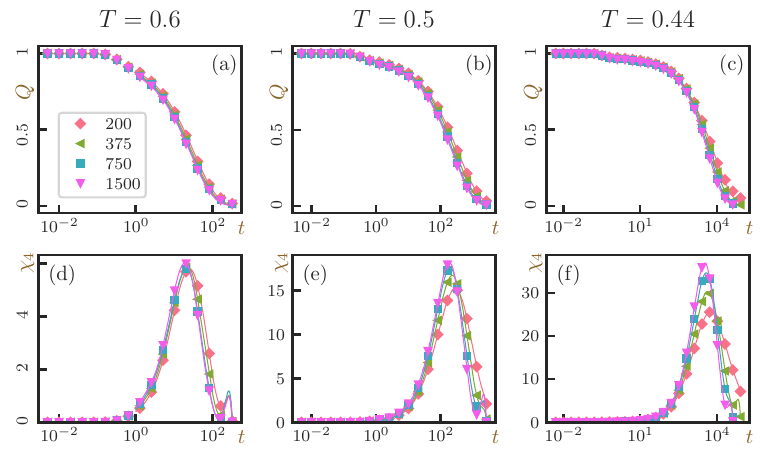}
\end{center}
\caption{
Semi-log plots of the overlap function $Q(t)$ and the dynamical susceptibility $\chi_4(t)$ as functions of time. 
Panels (a-c) in the top row show the results for $Q(t)$, with symbols representing the simulation data and lines corresponding to fits using the two-mode model given in Eq.~\ref{eq:2mode}.
Panels (d-f) in the bottom row show the results for $\chi_4(t)$, with symbols representing the simulation data and lines corresponding to cubic-spline interpolations.
Although small peaks sometimes appear at long times in the interpolated lines, they are artifacts of the interpolation and do not affect the extraction of the main peak.
From left to right, the results correspond to 
$T = 0.6 (\approx T_{\rm ava}), 0.5, 0.44 (\approx T_{\rm MCT})$.
In all panels, different symbols represent different system sizes, as indicated in the legend of panel (a).
}
\label{fig:Q_chi}
\end{figure*}

\subsection{Avalanche criticality seen in dynamical susceptibility}\label{sec:chi4}
In the bottom row (d-f) of Fig.~\ref{fig:Q_chi}, we show the time evolution of dynamical susceptibility $\chi_4(t)$ at temperatures $T=0.6$, $0.5$, and $0.44$. 
Different symbols correspond to four system sizes, $N=200$, $375$, $750$, and $1500$, as indicated in the legend.
At the relatively high temperature $T=0.6$, finite-size effects are hardly observed (Fig.~\ref{fig:Q_chi}d). 
At $T=0.5$, the smallest system size, $N=200$, shows a clear deviation from the results of larger system sizes (Fig.~\ref{fig:Q_chi}e).
At a lower temperature very close to the MCT point, $T=0.44$, a noticeable deviation from the largest system size, $N=1500$, is also observed for $N=375$ and $750$ (Fig.~\ref{fig:Q_chi}f).

To discuss these finite-size effects more quantitatively and comprehensively, we measured the peak height, $\chi_4^\ast$, and the peak time, $\tau_4$, as functions of temperature $T$ and system size $N$.
Both $\chi_4^\ast$ and $\tau_4$ were estimated from cubic-spline interpolations of $\chi_4(t)$ measured using logarithmic time sampling.
In this subsection and Sec.~\ref{sec:vdos}, we present part of the same data as those in our accompanying Letter~\cite{DHLetter} to make comparison with other observables and the results in previous works.

\begin{figure*}[t]
\begin{center}
\includegraphics[width=.8\linewidth,angle=0]{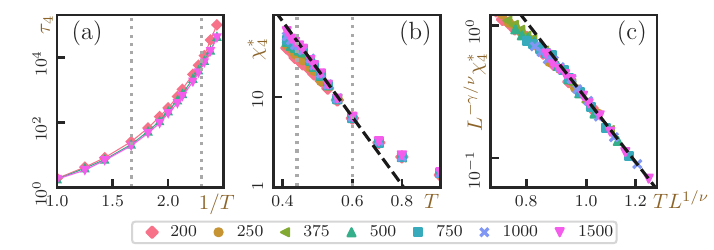}
\end{center}
\caption{
(a) Semi-log plot of the peak time, $\tau_4$, of the dynamical susceptibility $\chi_4(t)$ as a function of inverse temperature. For clarity, only the results for $N=200$, $500$, and $1500$ are shown.
(b) Log-log plot of the peak value of the dynamical susceptibility, $\chi_4^\ast$, as a function of temperature.
(c) Finite-size scaling of $\chi_4^\ast$ using the critical exponents obtained in the main text, $\nu \approx 3.2$ and $\gamma \approx 6.0$. 
Only the results for $T \le T_{\rm ava}=0.6$ are shown.
In panels (a) and (b), the vertical dotted lines indicate the positions of the temperatures $T_{\rm ava} \approx 0.6$ and $T_{\rm MCT} \approx 0.435$.
In panels (b) and (c), the dashed lines represent the power-law behavior $\chi_4^\ast \sim T^{-\gamma}$.
In all panels, different symbols represent different system sizes, as indicated in the legends below the panels.}
\label{fig:chi4star}
\end{figure*}

As shown in Fig.~\ref{fig:chi4star}(a), the system-size dependence of $\tau_4$ reaches roughly a factor of 2.5 at the lowest temperature. 
Compared with the very strong temperature dependence, which spans more than five orders of magnitude, the system-size dependence observed here is relatively modest.
In contrast, as shown in Fig.~\ref{fig:chi4star}(b), $\chi_4^\ast$ exhibits significant finite-size effects in the low-temperature regime $T \le T_{\rm ava} \approx 0.6$.
In the accompanying Letter\cite{DHLetter}, we showed that the finite-size effects of $\chi_4^\ast$ can be explained by the avalanche criticality picture proposed in Ref.~\cite{tahaeiScalingDescriptionDynamical2023a}, which was introduced using the T-EPM. 
Below, we recap the main points of our discussion in the Letter.

In the Letter, following Ref.~\cite{tahaeiScalingDescriptionDynamical2023a}, we introduced a scaling ansatz based on zero-temperature avalanche criticality:
\begin{align}
\xi &\sim T^{-\nu}, \label{eq:ansatz1} \\
\chi_4^\ast &\sim T^{-\gamma}. \label{eq:ansatz2}
\end{align}
Here, we introduced the critical correlation length of avalanches, $\xi$ (its physical interpretation is provided in Sec.~\ref{sec:xi_max}), along with the critical exponents $\gamma$ and $\nu$.
Eqs.~\ref{eq:ansatz1} and \ref{eq:ansatz2} indicate that the critical correlation length $\xi$ increases as the temperature decreases, which in turn leads to an increase in the dynamical susceptibility.
The fractal dimension $d_f$ can also be defined via the relation between $\xi$ and $\chi_4^\ast$:
\begin{align}
\chi_4^\ast \sim \xi^{d_f}, \label{eq:ansatz3}
\end{align}
and combining Eqs.~(\ref{eq:ansatz1})-(\ref{eq:ansatz3}) yields the scaling relation
\begin{align}
\nu d_f = \gamma.
\end{align}

We then showed that by adopting the conventionally well-established dynamic correlation length~\cite{karmakarGrowingLengthTime2009a,karmakarAnalysisDynamicHeterogeneity2010,tahGlassTransitionSupercooled2018a} as the critical correlation length $\xi$, the finite-size scaling of $\chi_4^\ast$ can be achieved with high accuracy.
The critical exponent $\nu$ was determined as $\nu \approx 3.2$ from a power-law fit of $\xi$ in the low-temperature regime $T \le T_{\rm ava} \approx 0.6$. 
As discussed later in Sec.~\ref{sec:vdos}, this temperature range is justified as the regime where critical behavior is expected, based on measurements of the vibrational density of states of inherent structures.

The critical exponents $\gamma$ and $d_f$ were determined using the method we developed for analyzing avalanche criticality in the plasticity of sheared glasses~\cite{oyamaInstantaneousNormalModes2021b,oyamaScaleSeparationShearInduced2024a}.\footnote{Previous works~\cite{oyamaInstantaneousNormalModes2021b,oyamaScaleSeparationShearInduced2024a} focused on zero-temperature systems and analyzed instantaneous normal modes. 
In the present study, where finite temperatures are considered, we focus on the normal modes of saddle configurations to remove effects of apparent instabilities arising from thermal fluctuations.
Notice that normal modes are the eigenmodes of the dynamical matrix.
Within the current setup, the dynamical matirix is merely identical to the Hessian matrix of the total potential energy of the system with respect to particle positions. 
}
In this method, we focus on the number of negative eigenvalues, $N^\dagger_{\rm saddle}$, of the dynamical matrix at a saddle configuration obtained from an instantaneous equilibrium snapshot (for technical details on how the saddle configurations are obtained, see Sec.~\ref{sec:saddle}).
In this analysis, the number of negative eigenvalues (corresponding to unstable eigenmodes) is considered to correspond to the number of plastic rearrangement events, i.e., STZs.
Within the avalanche criticality picture, which posits that these STZs form avalanche-like spatiotemporal correlations, one can show the relation $N^\dagger_{\rm saddle} \sim N T^{\nu d_f - \gamma}$, based on general phenomenological arguments\cite{DHLetter}.
Consistent with this argument, the fraction of unstable modes,
$f^\dagger_{\rm saddle} \equiv \frac{N^\dagger_{\rm saddle}}{dN}$, shows no system-size dependence and follows a power-law behavior, $f^\dagger_{\rm saddle} \sim T^{3.6}$, again, in the critical regime $T\le T_{\rm ava}$.
Together with the previously obtained value $\nu \approx 3.2$, we find $\gamma \approx 6.0$ and $d_f \approx 1.9$.
The finite-size scaling results using these values of $\nu$ and $\gamma$ are shown in Fig.~\ref{fig:chi4star}c.
The high-quality scaling collapse demonstrates the validity of the avalanche criticality picture and the values of the critical exponents.
We emphasize that these exponents were determined independently of $\chi_4^\ast$, from measurements of other observables, $\xi$ and $f^\dagger_{\rm saddle}$, and from phenomenologically derived scaling relations.
Therefore, we interpreted the temperature and system-size dependence of $\chi_4^\ast$ as being indeed described by criticality with $T=0$ as the critical point, in agreement with the arguments of Ref.~\cite{tahaeiScalingDescriptionDynamical2023a}.

We note that, it has been reported that the dynamical susceptibility $\chi_4^\ast$ saturates below $T_{\rm MCT}$ in refs.~\cite{coslovichDynamicThermodynamicCrossover2018a,dasCrossoverDynamicsKobAndersen2022a}, although such a crossover effect is not clearly detectable within the temperature range investigated in the present study.
The saturation of $\chi_4^\ast$ suggests that the criticality regime has also a lower bound, as $T_{\rm MCT}\le T\le T_{\rm ava}$.
The meaning of this lower bound at around $T_{\rm MCT}$ will be discussed in detail in Sec.~\ref{sec:xi_max} in relation to another important phenomenon, the localization of the unstable saddle modes, which will be investigated in Sec.~\ref{sec:deloc}.

\subsection{Overlap function and two-mode relaxation model}\label{sec:Q}
In this section, we present measurement results for the $T$ and $N$ dependence of the relaxation function $Q(t)$, which quantifies the average relaxation behavior. 
We further demonstrate that such parameter dependence of $Q(t)$ (in particular, that of the stretching exponent characterizing its relaxation) can be consistently interpreted within the avalanche criticality picture.

Figures~\ref{fig:Q_chi}(a)-(c) in the upper row show the results for $Q(t)$ at temperatures $T=0.6$, $0.5$, and $0.44$.
As in the case of $\chi_4$, different symbols distinguish the results for system sizes $N=200$, $375$, $750$, and $1500$.
As the temperature is lowered, finite-size effects become more pronounced, and the relaxation becomes slower for smaller system sizes.
This behavior is consistent with the results reported in Ref.~\cite{karmakarGrowingLengthTime2009a}.
Now we perform a more comprehensive analysis of the parameter dependence observed in Figs.~\ref{fig:Q_chi}(a)-(c), using a two-mode relaxation model:
\begin{align}
    Q_{\rm 2m}(t)\equiv(1-f_{\rm c})\exp(-(t/t_{\rm s}))^2+f_{\rm c}\exp(-t/\tau_\alpha)^{\beta_{\rm KWW}}.\label{eq:2mode}
\end{align}
The first term on the right-hand side of Eq.~\ref{eq:2mode} represents the fast relaxation mode.
For the fast mode, following Ref.~\cite{dasCrossoverDynamicsKobAndersen2022a}, we adopt a Gaussian approximation and fix the exponent to 2.
The second term represents the slow relaxation mode, for which we assume a Kohlrausch-Williams-Watts (KWW)-type stretched exponential function~\cite{williamsNonsymmetricalDielectricRelaxation1970}.
This two-mode model, Eq.~\ref{eq:2mode}, involves four parameters: the fast-mode timescale $t_{\rm s}$, the slow-mode weight $f_{\rm c}$, the slow-mode relaxation time $\tau_\alpha$, and the stretching exponent $\beta_{\rm KWW}$, which characterizes the shape of the slow-mode relaxation function.
The overlap function $Q(t)$ obtained from molecular dynamics simulations was fitted using Eq.~\ref{eq:2mode}, and the four parameters were determined as functions of $T$ and $N$.
Since the present study focuses on slow dynamics, we discuss in the main text only the results for the two key parameters characterizing the slow mode, namely the relaxation time $\tau_\alpha$ and the stretching exponent $\beta_{\rm KWW}$.
The results for the remaining two parameters, $f_{\rm c}$ and $t_{\rm s}$, which quantify the contribution of the fast mode, are presented in Appendix~\ref{ap:fast}.
Although Eq.~\ref{eq:2mode} is an empirical expression, it not only provides an excellent fit to the simulation data but also allows for a meaningful discussion of the $T$- and $N$-dependent slow dynamics in terms of the model parameters, as demonstrated below.

\begin{figure}[t]
\begin{center}
\includegraphics[width=\linewidth,angle=0]{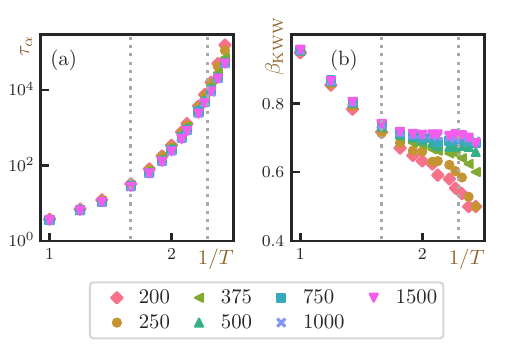}
\end{center}
\caption{
Slow-mode parameters obtained from fitting $Q(t)$ using the two-mode model, Eq.~\ref{eq:2mode}.
(a) Semi-log plot of the $\alpha$-relaxation time $\tau_\alpha$ as a function of inverse temperature.
(b) Linear plot of the stretching exponent $\beta_{\rm KWW}$ as a function of inverse temperature.
In both panels, the vertical dashed lines indicate the locations of $T_{\rm ava}\approx 0.6$ and $T_{\rm MCT}\approx 0.435$, and different symbols represent different system sizes as indicated in the legend below the panels.
}
\label{fig:2mode}
\end{figure}

\begin{figure*}[t]
\begin{center}
\includegraphics[width=.8\linewidth,angle=0]{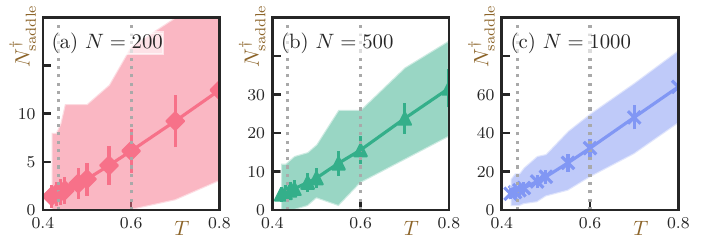}
\end{center}
\caption{
Linear plot of the number of unstable modes at saddle-point configurations, $N_{\rm saddle}^\dagger$, as a function of temperature.
Symbols indicate the mean values, error bars represent the standard deviations, and the shaded regions depict the range between the minimum and maximum values.
Vertical dashed lines indicate the locations of $T_{\rm ava}\approx 0.6$ and $T_{\rm MCT}\approx 0.435$.
Results are shown for (a) $N=200$, (b) $N=500$, and (c) $N=1000$.
}
\label{fig:Ndagger}
\end{figure*}

Figure~\ref{fig:2mode} shows the $T-$ and $N-$ dependence of the slow-mode parameters $\tau_\alpha$ and $\beta_{\rm KWW}$.
As shown in Fig.~\ref{fig:2mode}(a), similarly to the case of $\tau_4$, the system-size dependence of $\tau_\alpha$ is relatively weak compared to its temperature dependence.
Note that previous studies have shown a crossover in the $1/T$ dependence of $\tau_\alpha$ from fragile to strong behavior around the MCT point~\cite{coslovichDynamicThermodynamicCrossover2018a,dasCrossoverDynamicsKobAndersen2022a}.
A consistent crossover behavior is observed in our results in Fig.~\ref{fig:2mode}(a).

Figure~\ref{fig:2mode}(b) shows the dependence of $\beta_{\rm KWW}$ on $T$ and $N$.
For $N \ge 750$, $\beta_{\rm KWW}$ is nearly constant in the temperature range
$T_{\rm ava} \ge T \ge T_{\rm MCT}$, where $\chi_4^\ast$ follows avalanche criticality.
This plateau of $\beta_{\rm KWW}$ can be interpreted from the perspective of avalanche criticality based on a phenomenological picture explained below.
Within our avalanche criticality picture, structural relaxation proceeds through the occurrence of avalanches, and therefore the relaxation time is described by the distribution of waiting times between avalanches, $P(\tau_{\rm ava})$. 
Here, $\tau_{\rm ava}$ denotes the waiting time between successive avalanches and, in the presence of avalanche criticality, the distribution
$P(\tau_{\rm ava}/\langle\tau_{\rm ava}\rangle)$, normalized by the mean value
$\langle\tau_{\rm ava}\rangle$, is expected to collapse onto a single master curve independent of the control parameter (here, temperature)~\cite{corralLongTermClusteringScaling2004a,janicevicThresholdinducedCorrelationsRandom2018a,zhangScalingSlipAvalanches2017,liDoublePowerlawUniversal2024a}.
On the other hand, Ref.~\cite{palmerModelsHierarchicallyConstrained1984} discussed that a stretched exponential function can be expressed as a superposition of exponential relaxation modes with various time scales (hereafter, we refer to this picture as the multiple relaxation modes picture):
\begin{align}
\exp(-t/\tau_\alpha)^{\beta_{\rm KWW}}=\int d\tau\exp(-t/\tau)P(\tau).\label{eq:multi_mode}
\end{align}
Taking this picture into account, the same value of $\beta_{\rm KWW}$ is obtained as long as
$P(\tau/\langle\tau\rangle)$ is identical (see Appendix~\ref{ap:multi} for a detailed explanation).
Therefore, assuming that $P(\tau/\langle\tau\rangle)$ becomes identical when $P(\tau_{\rm ava}/\langle\tau_{\rm ava}\rangle)$ is identical, the multiple relaxation modes picture implies that, in the presence of avalanche criticality, $\beta_{\rm KWW}$ is independent of temperature.
The converse is not necessarily true, and the specific relationship between
$P(\tau)$ and $P(\tau_{\rm ava})$ remains unknown.
Nevertheless, based on the discussion in this paragraph, we regard the plateau of
$\beta_{\rm KWW}$ observed in the critical regime $T_{\rm MCT}\le T \le T_{\rm ava}$ as supporting evidence for the validity of the avalanche criticality hypothesis.
We note that the multiple relaxation mode picture Eq.~\ref{eq:multi_mode} is a simplified description, and it remains a matter of debate whether it accurately reflects the underlying microscopic dynamical processes~\cite{shangLocalGlobalStretched2019,berthierSelfInducedHeterogeneityDeeply2021}.
For example, relaxation along a single trajectory is already composed of multiple local rearrangements and exhibits stretched-exponential behavior.
In such a case, it is not obvious whether a picture based on a superposition of exponential functions, as in Eq.~\ref{eq:multi_mode}, is appropriate.
See Appendix~\ref{ap:multi} for a more detailed discussion of this multiple relaxation modes picture.

On the other hand, for smaller system sizes ($N \le 500$), $\beta_{\rm KWW}$ takes smaller values due to finite-size effects even in the scaling regime $T_{\rm MCT}\le T\le T_{\rm ava}$.
According to the multiple relaxation modes picture Eq.~\ref{eq:multi_mode}, a reduction of
$\beta_{\rm KWW}$ suggests a broadening of the distribution $P(\tau)$.
Moreover, based on our picture in which relaxation proceeds through avalanches, it is expected that we can quantitatively discuss  such system-size dependence of $P(\tau)$ in terms of our quantitative measure of avalanches, $N^\dagger_{\rm saddle}$ (defined as the number of negative eigenvalues of a saddle configuration: see Sec.~\ref{sec:chi4}).
In the following, we demonstrate the validity of this expectation.
Figure~\ref{fig:Ndagger} shows the temperature dependence of $N^\dagger_{\rm saddle}$ for $N=200$, $500$, and $1000$.
In particular, we present not only the average and the standard deviation, but also the maximum and minimum values.
From this figure, we can tell that for $N=1000$, unstable modes are always present at all temperatures investigated in the present study, as indicated by the fact that the minimum value of $N^\dagger_{\rm saddle}$ remains positive.
In contrast, for $N=200$ and $500$, unstable modes are not always present at $T \le 0.6$ and $0.45$, respectively, as indicated by the minimum value of $N^\dagger_{\rm saddle}$ becoming zero.
These temperature ranges roughly correspond~\footnote{The correspondence is not perfect. The slight difference between the two ranges is likely due to entropy effects discussed in Secs.~\ref{sec:t_avalanches} and \ref{sec:PEL_previous}.} to the temperature ranges shown in Fig.~\ref{fig:2mode}(b), in which finite-size effects in $\beta_{\rm KWW}$ become non-negligible for $N=200$ and $500$.
This suggests that the finite-size effects arise from the breakdown of self-similarity associated with the appearance of configurations with $N^\dagger_{\rm saddle}=0$.
From the viewpoint of the relaxation-time distribution $P(\tau)$, this indicates that the long relaxation times linked to fully stable configurations modifies the value of $\beta_{\rm KWW}$.

Based on the results shown in Fig.~\ref{fig:2mode}b and the discussions in this section, we consider that the parameter dependence of the stretched exponent $\beta_{\rm KWW}$, a key parameter characterizing relaxational dynamics, can also be understood from the perspective of avalanche criticality.

\begin{figure}[t]
\begin{center}
\includegraphics[width=.8\linewidth,angle=0]{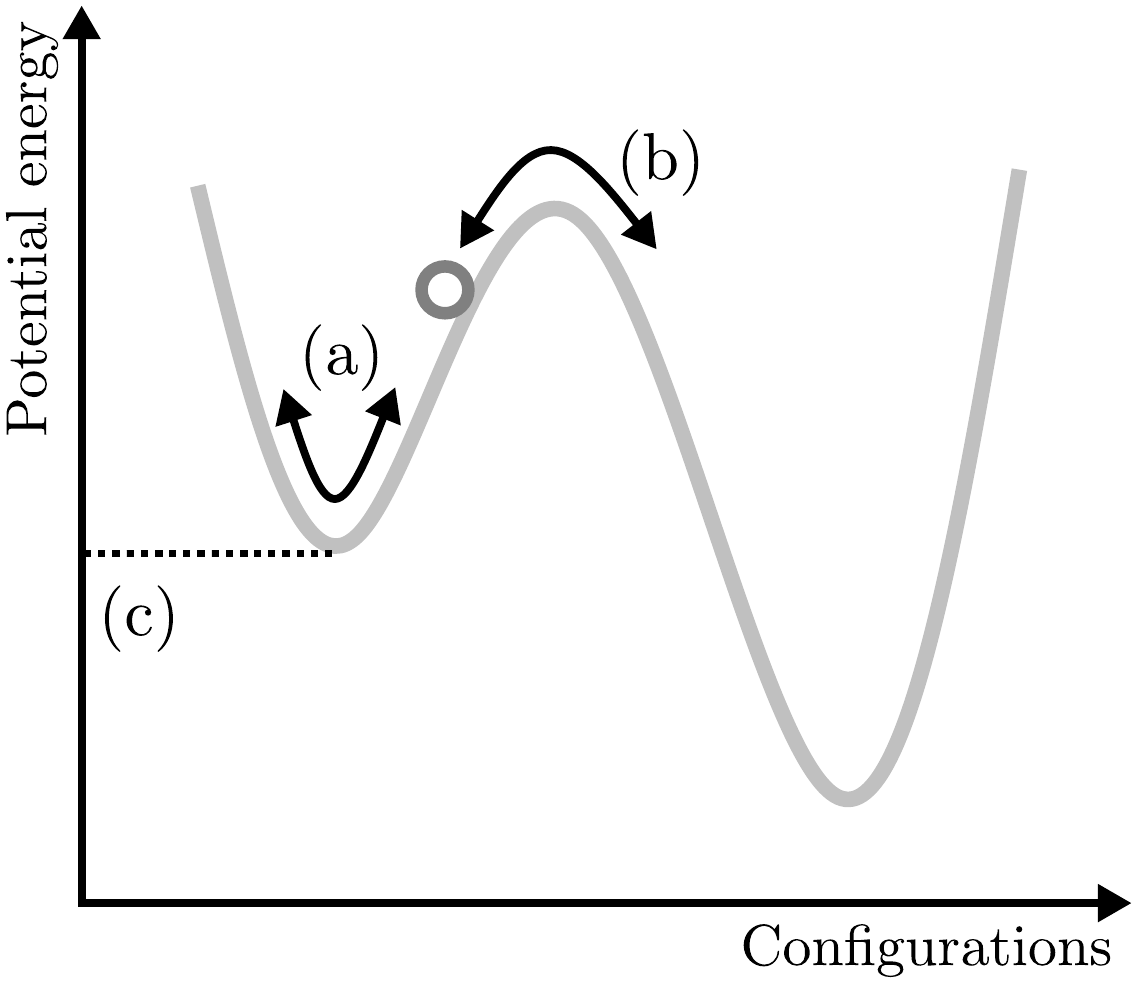}
\end{center}
\caption{
Schematic illustration of the potential energy landscape.
The vertical axis represents the potential energy, while the horizontal axis denotes particle configurations.
The white circle indicates an instantaneous state of the system.
(a) The inherent structure corresponding to the system state, (b) a saddle point (the nearest stationary point), and (c) the energy level of the inherent structure.
For simplicity, the PEL is shown as a one-dimensional function; in reality, it is defined on a high-dimensional hypersurface of dimension $dN-d$.}
\label{fig:PEL}
\end{figure}

\section{Results: potential energy landscape}\label{sec:results2}
In the previous section, we proposed a unified interpretation of the dependence of dynamical observables such as $Q(t)$ and $\chi_4(t)$ on $T$ and $N$ in terms of avalanche criticality.
The fact that the avalanche critical exponent $\gamma$ can be derived from an argument based on the number of unstable modes at saddle-point configurations, $N_{\rm saddle}^\dagger$, suggests that the dynamics may be governed by the PEL schematically illustrated in Fig.~\ref{fig:PEL}.
Indeed, the importance of the PEL in governing the dynamics of supercooled liquids has been discussed in many previous studies
(The relation between these earlier works and the present study is discussed in Sec.~\ref{sec:ava_PEL}.).

In this section, we examine the temperature dependence of the properties of the PEL from three different perspectives, namely, the vibrational density of states of inherent structures, the localization transition of unstable modes at sadle configurations, and the energy levels of inherent structures.

\subsection{Vibrational density of states of inherent structures}\label{sec:vdos}

\begin{figure*}[t]
\begin{center}
\includegraphics[width=.8\linewidth,angle=0]{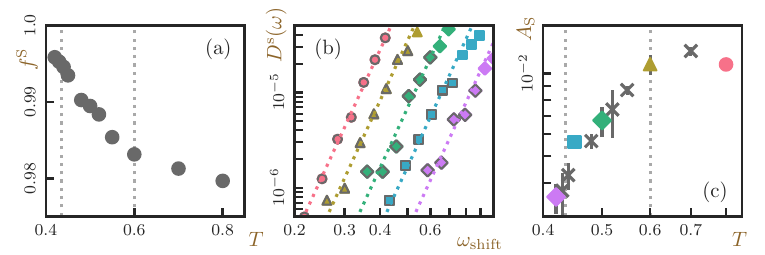}
\end{center}
\caption{
Normal-mode analysis of inherent structures obtained at different parent temperatures for the system with $N=1000$.
For each temperature, statistical quantities were computed using 40960 independent equilibrium configurations.
(a) Linear plot of the temperature dependence of the fraction of stable samples, $f^{\rm S}$.
(b) Log-log plot of the low-frequency region of the vibrational density of states calculated from stable configurations only, $D^{\rm S}(\omega)$.
Different symbols correspond to the temperatures indicated by the same symbols in panel (c).
For better visual clarity, the data are plotted as a function of a shifted frequency, $\omega_{\rm shift}$, where the shift along the frequency axis is chosen separately for each temperature (see Appendix~\ref{ap:dos} for results plotted as a function of the unshifted frequency $\omega$).
Dotted lines indicate fits of the data to an $\omega^{6.5}$ power law at each temperature.
(c) Log-log plot of the temperature dependence of the coefficient $A_{\rm S}$ obtained from fitting $D^{\rm S}(\omega)$ to the $\omega^{6.5}$ law.
Error bars represent the fitting uncertainties estimated from the covariance matrix obtained by least-squares fits.
In panels (a) and (c), vertical dashed lines indicate the positions of $T_{\rm ava}=0.6$ and $T_{\rm MCT}=0.435$.
}
\label{fig:vdos}
\end{figure*}

As the first quantitative measure of the properties of the PEL, we examine the vibrational density of states of inherent structures.
An inherent structure is defined as a configuration obtained by minimizing the potential energy starting from an arbitrary initial configuration.
And the vibrational density of states is defined as the probability distribution of the square roots of the eigenvalues, i.e., the eigenfrequencies, of the dynamical matrix.
Accordingly, the vibrational density of states of inherent structures provide the statistical information of (the square roots of) the local curvatures associated with a single basin of the PEL to which the analyzed configuration belongs, as schematically illustrated in Fig.~\ref{fig:PEL}(a).
In particular, the low-frequency limit of the vibrational density of states, $D_0(\omega)$, is known to exhibit a universal $D_0(\omega)=A_4\omega^4$ scaling law, often referred to as the non-Debye law, across a wide variety of amorphous solids~\cite{lernerStatisticsPropertiesLowFrequency2016,mizunoContinuumLimitVibrational2017a,shimadaAnomalousVibrationalProperties2018a,kapteijnsUniversalNonphononicDensity2018,bonfantiUniversalLowFrequencyVibrational2020,richardUniversalityNonphononicVibrational2020}.
Here, $\omega$ stands for the eigenfrequency.
It is known that the non-Debye law originates from nonphononic, quasilocalized vibrational modes (QLVs) associated with local plastic rearrangements, i.e., structural instabilities~\cite{maloneyUniversalBreakdownElasticity2004,gartnerNonlinearPlasticModes2016}.
Therefore, the coefficient $A_4$ in the non-Debye law can be regarded as an indicator of the stability of the system and provides important information for discussing zero-temperature avalanche criticality.
Moreover, it is known that when we investigate the inherent structures of equilibrium configurations, the resulting statistical properties exhibit a clear parent-temperature dependence.
For example, in Ref.~\cite{wangLowfrequencyVibrationalModes2019a}, the parent-temperature dependence of $D_0(\omega)$ was investigated for a polydisperse system~\cite{ninarelloModelsAlgorithmsNext2017a} (we call this polydisperse system introduced in ref.~\cite{ninarelloModelsAlgorithmsNext2017a} the swap system hereafter).
In Ref.~\cite{wangLowfrequencyVibrationalModes2019a}, it was reported that for sufficiently large system sizes ($N \ge 48{,}000$), the non-Debye law $D_0(\omega)=A_4 \omega^4$ is observed independently of parent temperature, with the coefficient $A_4$ exhibiting a clear temperature dependence.  
Accordingly, $A_4$ serves as an indicator of the parent-temperature-dependent stability of the system.

In this section, we present our results for the non-Debye law in the KAM.
In the accompanying Letter, we showed that this analysis allows us to identify the temperature range in which zero-temperature criticality is at play.
In the present paper, in order to discuss the relation between this analysis and other observables, we present part of the same data.
Inherent structures were obtained by minimizing the potential energy using the FIRE algorithm~\cite{bitzekStructuralRelaxationMade2006a,guenoleAssessmentOptimizationFast2020a}.

\subsubsection{Vibrational density of states of stable configurations}\label{sec:vdos_stable}
As discussed above, for amorphous solids with sufficiently large system sizes, the low-frequency limit of the vibrational density of states follows a power-law behavior $D_0(\omega)\sim \omega^\alpha$, with a universal exponent $\alpha=4$~\cite{lernerStatisticsPropertiesLowFrequency2016,mizunoContinuumLimitVibrational2017a,shimadaAnomalousVibrationalProperties2018a,kapteijnsUniversalNonphononicDensity2018,bonfantiUniversalLowFrequencyVibrational2020,richardUniversalityNonphononicVibrational2020}. However, for relatively small system sizes such as those considered in the present study ($N\le 1500$), it is known that finite-size effects can cause the measured exponent $\alpha$ to depend on the parent temperature~\cite{lernerFinitesizeEffectsNonphononic2020,xuLowfrequencyVibrationalDensity2024b}.

On the other hand, a recent study has shown that a universal power-law behavior can be extracted even for small system sizes by classifying samples based on an extended Hessian matrix that incorporates boundary-condition degrees of freedom~\cite{xuLowfrequencyVibrationalDensity2024b}.
In Ref.~\cite{xuLowfrequencyVibrationalDensity2024b}, a method was proposed to classify a given configuration as either stable or unstable depending on whether the extended Hessian possesses negative eigenvalues.
It was reported that the low-frequency limit of the vibrational density of states obtained from an ensemble consisting only of stable samples, $D_0^{\rm S}(\omega)$, follows a universal power-law behavior $D_0^{\rm S}(\omega)=A_{\rm S}\omega^{6.5}$ even for small system sizes.
In the present study, we perform the same analysis for simulation data of the KAM.

We first show in Fig.~\ref{fig:vdos}(a) the temperature dependence of the fraction of stable samples, $f^{\rm S}$, for the system with $N=1000$.
Over the entire temperature range investigated, more than 97\% of the samples are classified as stable, and, consistent with the results of Ref.~\cite{xuLowfrequencyVibrationalDensity2024b}, the fraction of stable samples increases upon cooling.
In particular, $f^{\rm S}$ appears to exhibit an inflection point near the MCT transition temperature.
While the existence of an inflection point itself is reasonable given that $f^{\rm S}$ cannot exceed unity even in the low-temperature limit, its location in the vicinity of the MCT point is intriguing.
The vibrational density of states measured for the ensemble consisting only of stable samples, $D^{\rm S}(\omega)$, is shown in Fig.~\ref{fig:vdos}(b) for several parent temperatures, $T=0.8, 0.6, 0.5, 0.48,$ and $0.42$.
Indeed, for all temperatures studied, the low-frequency limit is found to be consistent with the scaling form $D_0^{\rm S}(\omega)=A_{\rm S}\omega^{6.5}$.

\subsubsection{Critical temperature regime}\label{sec:critical_regime}
As shown in the previous subsection, consistent with the report of Ref.~\cite{xuLowfrequencyVibrationalDensity2024b}, we confirmed that the low-frequency limit of the vibrational density of states of stable samples, $D_0^{\rm S}(\omega)$, can be described by the functional form $D_0^{\rm S}(\omega)=A_{\rm S}\omega^{6.5}$.
This implies that the temperature dependence of $D_0^{\rm S}(\omega)$ can be quantitatively characterized solely by the magnitude of the coefficient $A_{\rm S}$.
Accordingly, in Fig.~\ref{fig:vdos}(c) we plot the coefficient $A_{\rm S}$ as a function of temperature.
We find that $A_{\rm S}$ remains approximately constant above a threshold temperature $T_{\rm ava}\approx 0.6$, while it decreases monotonically below $T_{\rm ava}$.
A reduction in $A_{\rm S}$ corresponds to a decrease in the number of QLVs, which act as destabilizing modes that trigger plastic rearrangements.
Therefore, the decrease in $A_{\rm S}$ implies that the system becomes more stable upon cooling below $T_{\rm ava}$.

The decrease of $A_{\rm S}$ below $T_{\rm ava}$ is qualitatively consistent with the change in stability reported by Tahaei et al. in Ref.~\cite{tahaeiScalingDescriptionDynamical2023a}.
Using a lattice-based model, the T-EPM, Tahaei et al. proposed an avalanche criticality scenario for thermal relaxation process of supercooled liquids.
They also measured the probability distribution of a stability parameter $x$, which corresponds to the distance from local yielding at each lattice site (see Sec.~\ref{sec:t_avalanches} for details of the thermal EPM), and reported that the stability increases upon cooling.
Their results imply that there exists a correlation between such temperature-dependent stability changes and avalanche criticality.
As discussed in Sec.~\ref{sec:results1}, our analysis of $\chi_4^\ast$ and $\beta_{\rm KWW}$ indicates that, in the KAM, avalanche criticality is realized in the temperature range $T \le T_{\rm ava}\approx 0.6$.
Notably, Fig.~\ref{fig:vdos}(c) reveals that, within this temperature range, the system exhibits enhanced stability upon cooling, consistent with the behavior reported in Ref.~\cite{tahaeiScalingDescriptionDynamical2023a}. 
We interpret this correspondence as indirect support for the avalanche criticality picture.
On the other hand, as discussed in Sec.~\ref{sec:xi_max}, it is known that the peak value of the dynamical susceptibility $\chi_4^\ast$ (and thus the dynamical correlation length $\xi$ as well) tends to saturate in the vicinity of $T_{\rm MCT}$~\cite{coslovichDynamicThermodynamicCrossover2018a,dasCrossoverDynamicsKobAndersen2022a}.
A corresponding qualitative change is not clearly detected in $A_{\rm S}$.

\subsection{Localization of unstable modes at saddle configurations}\label{sec:deloc}
Next, we examine the properties of the saddle points of the PEL, schematically illustrated in Fig.~\ref{fig:PEL}(b).
As briefly recapped in Sec.~\ref{sec:chi4}, in the accompanying Letter~\cite{DHLetter} we successfully derived the avalanche critical exponent $\gamma$ by focusing on the number of unstable modes (represented by negative eigenvalues of dynamical matrix) associated with saddle-point configurations.
While that analysis considered only the eigenvalues, in the present section we investigate the properties of the corresponding eigenvectors, following the analysis of Ref.~\cite{coslovichLocalizationTransitionUnderlies2019b}.

\subsubsection{Saddle-point configurations}\label{sec:saddle}
In Ref.~\cite{coslovichLocalizationTransitionUnderlies2019b}, the degree of spatial localization of unstable modes at saddle-point configurations was systematically investigated.
A saddle-point configuration corresponds to an unstable stationary point of the PEL, as schematically illustrated in Fig.~\ref{fig:PEL}(b), and the unstable modes of the associated dynamical matrix  reflect the local curvatures at the top of energy barriers in the PEL.
Note that in Fig.~\ref{fig:PEL}(b), the horizontal axis is shown as one-dimensional for visualization purposes.
In reality, however, the PEL is defined on a high-dimensional hypersurface with $dN-d$ degrees of freedom, and at a saddle-point configuration the system is unstable along some directions while being stable along the majority of the remaining ones~\cite{coslovichLocalizationTransitionUnderlies2019b,angelaniSaddlesEnergyLandscape2000a,broderixEnergyLandscapeLennardJones2000b}.
In the present study, saddle-point configurations were obtained by minimizing $W\equiv \frac{1}{N}\left(\boldsymbol{\nabla}U\right)^2$ using the FIRE algorithm~\cite{bitzekStructuralRelaxationMade2006a,guenoleAssessmentOptimizationFast2020a}\footnote{When using simple gradient-based methods to obtain saddle-point configurations, the algorithm may occasionally converge to so-called quasi-saddles, i.e., local solutions that correspond to inflection points of the PEL~\cite{coslovichLocalizationTransitionUnderlies2019b}.
We eliminate quasi-saddle modes by introducing a threshold $\lambda_{\rm th}=-10^{-4}$, following Ref.~\cite{coslovichLocalizationTransitionUnderlies2019b}.}.
Here, $U$ denotes the total potential energy of the system.
In the following, we first briefly review the analysis presented in Ref.~\cite{coslovichLocalizationTransitionUnderlies2019b}, and then report the results of applying the same analysis to the KAM system.

\begin{figure*}[!tb]
\begin{center}
\includegraphics[width=\linewidth,angle=0]{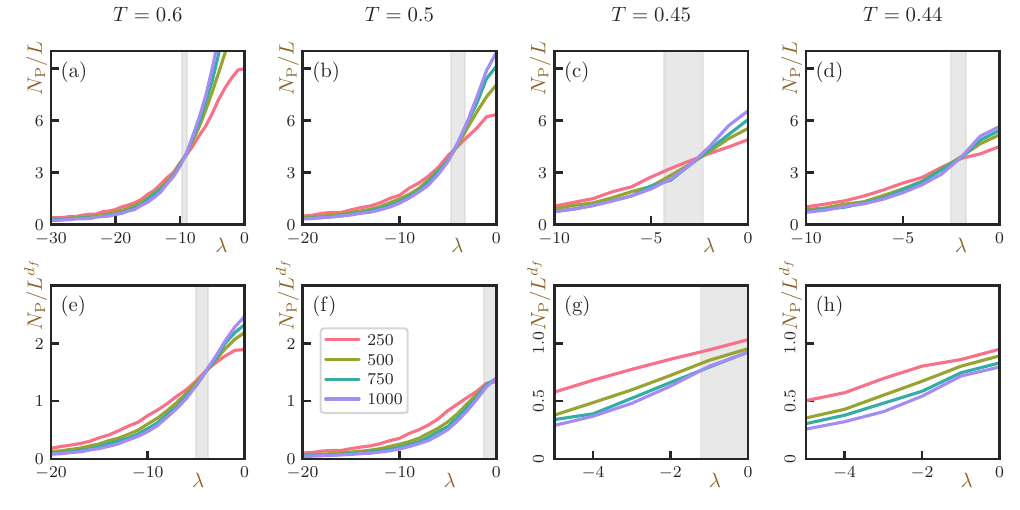}
\end{center}
\caption{
(a-d) Linear plots of $N_{\rm p}/L$, the number of particles contributing to each mode scaled by the system linear dimension $L$, shown as a function of the eigenvalue $\lambda$.
(e-h) Linear plots of $N_{\rm p}/L^{d_f}$, the number of particles contributing to each mode scaled by $L^{d_f}$ under the assumption of fractal eigenvectors, shown as a function of $\lambda$.
In each row, the panels from left to right correspond to different temperatures ($T=0.6$, $0.5$, $0.45$, and $0.44$), while different curves represent different system sizes as indicated in the legend.
The shaded regions indicate the range between the minimum and maximum values of the mobility edge $|\lambda_e|$ detected at each temperature, defined as the crossing points of the results for different system sizes.
}
\label{fig:me}
\end{figure*}

\begin{figure}[t]
\begin{center}
\includegraphics[width=\linewidth,angle=0]{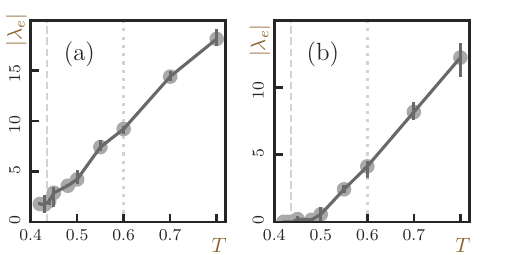}
\end{center}
\caption{
Linear plots showing the temperature dependence of the absolute value of the mobility edge estimated from the results in Fig.~\ref{fig:me}.
(a) Mobility edge $|\lambda_e|$ detected using $N_{\rm p}/L$.
(b) Mobility edge $|\lambda_e|$ detected using $N_{\rm p}/L^{d_f}$.
Circle symbols indicate the mean values of the crossing points identified in Fig.~\ref{fig:me}, while the endpoints of the error bars represent the minimum and maximum values.
}
\label{fig:me_t}
\end{figure}

\subsubsection{Analysis based on the linear-structure assumption}\label{sec:linear_me}
In Ref.~\cite{coslovichLocalizationTransitionUnderlies2019b}, the number of particles contributing to each eigenmode, $N_p$, defined by the following expression, was first measured as a function of the eigenvalue $\lambda$ at each temperature:
\begin{align}
N_p(\lambda) = \langle p_k\rangle_{\lambda_k=\lambda}.
\end{align}
Here, $p_k=\left(\sum_{i=1}^N|\boldsymbol{e}_{k,i}|^4\right)^{-1}$ denotes the (unnormalized) participation ratio of the eigenmode $\boldsymbol{e}_k$ (normalized as $|\boldsymbol{e}_k|=1$), $\boldsymbol{e}_{k,i}$ represents the $d$-dimensional component associated with particle $i$ of the $dN$-dimensional eigenvector $\boldsymbol{e}_k$, and $\langle\cdot\rangle_{\lambda_k=\lambda}$ indicates an average taken over the ensemble of modes with eigenvalue $\lambda_k=\lambda$.
The participation ratio $p_k$ provides a measure of how many particles effectively participate in a given eigenvector $\boldsymbol{e}_k$. It takes the value $p_k=N$ when all particles contribute equally, and $p_k=1$ when the mode is maximally localized on a single particle.
In Ref.~\cite{coslovichLocalizationTransitionUnderlies2019b}, the typical degree of spatial localization of modes associated with a given eigenvalue was analyzed by plotting $N_p(\lambda)/L$ as a function of $\lambda$ for different system sizes at fixed temperature.
If the typical mode at eigenvalue $\lambda$ has a spatially delocalized structure, the average number of particles contributing to the mode, $N_p(\lambda)$, grows at least linearly with the system linear dimension $L\sim N^{1/d}$.
In this case, $N_p(\lambda)/L$ increases with increasing $L$.
By contrast, if the modes are spatially localized, $N_p(\lambda)$ grows sublinearly with $L$, and consequently $N_p(\lambda)/L$ decreases as a function of $L$.
Therefore, if a localization-delocalization transition exists at a specific eigenvalue $\lambda_e$ (refered to as the mobility edge), one expects that $N_p(\lambda)/L$ for different $N$ intersect at this mobility edge $\lambda_e$.

In Ref.~\cite{coslovichLocalizationTransitionUnderlies2019b}, the $T$ and $N$ dependence of $N_p(\lambda)/L$ was investigated for several different systems with distinct setups, including different interparticle potentials and particle-size distributions.
As a result, it was found that, in all systems studied, at sufficiently high temperatures, there exists mobility edges $\lambda_e$ at which the results of different system sizes cross.
The temperature dependence of the absolute value $|\lambda_e|$ was then investigated and a very interesting behavior has been reported.
For nearly all systems examined, $|\lambda_e|$ was found to decrease upon cooling and become zero at a temperature slightly above the MCT point.
This implies that, below the MCT point, all unstable modes are spatially localized.
Note that mean-field theory predicts the disappearance of unstable modes themselves at the MCT transition~\cite{cavagnaRoleSaddlesMeanfield2001}.
Importantly, for the SiO$_2$ system modeled using the Coslovich-Pastore potential~\cite{coslovichDynamicsEnergyLandscape2009} (hereafter referred to as the CPP system), $|\lambda_e|$ remains finite even below the MCT transition temperature.
In Ref.~\cite{coslovichLocalizationTransitionUnderlies2019b}, the authors suggested that such qualitative differences might originate from differences in fragility.

In Figs.~\ref{fig:me}(a-d), following Ref.~\cite{coslovichLocalizationTransitionUnderlies2019b}, we plot $N_p/L$ measured for different system sizes $N$ as a function of the eigenvalue $\lambda$ for the KAM.
As indicated by the gray shaded regions in these figures, mobility edges $\lambda_e$ can be identified at all temperatures investigated.
That is, as shown in Fig.~\ref{fig:me_t}(a), the absolute value $|\lambda_e|$ does not vanish even below the MCT transition temperature at least within the temperature range explored in this study.
This result is qualitatively the same as that obtained for the CPP system in Ref.~\cite{coslovichLocalizationTransitionUnderlies2019b}.
However, the KAM is fragile for $T \ge T_{\rm MCT}$, whereas the CPP system is strong, indicating that the two systems differ qualitatively in their fragility.
This suggests that fragility does not play the dominant role in determining whether the mobility edge vanishes near $T_{\rm MCT}$.
Considering that, among the systems considered in Ref.~\cite{coslovichLocalizationTransitionUnderlies2019b}, the CPP system is the only one with attractive interactions, it is instead plausible that whether the interaction potential includes attractive forces plays a dominant role in this qualitative difference.

\subsubsection{Analysis based on the fractal-structure assumption}\label{sec:fractal_me}

The detection of the mobility edge $\lambda_e$ based on $N_{\rm p}/L$, as adopted in Ref.~\cite{coslovichLocalizationTransitionUnderlies2019b} and in the discussion of the previous subsection, implicitly assumes that delocalized structures are linear.
However, linear dependence on $L$ is only a necessary condition for an eigenmode to have a system-spanning, delocalized structure.
On the other hand, as we have shown in the accompanying Letter through the analysis based on the avalanche criticality picture (see the short summary in Sec.~\ref{sec:chi4}), long-time relaxational dynamics exhibits a fractal structure characterized by $d_f \approx 1.9$.
It is therefore possible that, even at the level of normal modes, the associated structures are characterized by a fractal dimension larger than unity.
We thus assume that the normal modes are characterized by the same fractal dimension $d_f$ as the nonlinear avalanche dynamics (the validity of this assumption is discussed in Sec.~\ref{sec:xi_max}).
In this case, $\lambda_e$ should be detected not from $N_{\rm p}/L$, but rather from $N_{\rm p}/L^{d_f}$.

Figures~\ref{fig:me}(e-h) show $N_{\rm p}/L^{d_f}$ plotted as a function of $\lambda$ for various system sizes $N$ at $T = 0.6$, $0.5$, $0.45$, and $0.44$.
In panels (e-g), corresponding to $T \ge 0.45$, $\lambda_e$ can be detected, as in Figs.~\ref{fig:me}(a-d).
By contrast, at temperatures $T \le 0.44$, the results for different $N$ do not intersect (Fig.~\ref{fig:me}(h)).
Thus, when the analysis is performed under the assumption of fractality, one can detect a qualitative change in the PEL such that $|\lambda_e| = 0$ at $T \approx T_{\rm MCT}$ (in particular, slightly above $T_{\rm MCT}$), as in many of the systems studied in Ref.~\cite{coslovichLocalizationTransitionUnderlies2019b} (Fig.~\ref{fig:me_t}(b)).
As already mentioned in Sec.~\ref{sec:linear_me}, this also suggests that, in systems with attractive interactions, the fractal character of the normal modes may become more pronounced.
The relationship between the localization properties investigated here and avalanche criticality is discussed in Sec.~\ref{sec:xi_max}.

\begin{figure*}[!tb]
\begin{center}
\includegraphics[width=\linewidth,angle=0]{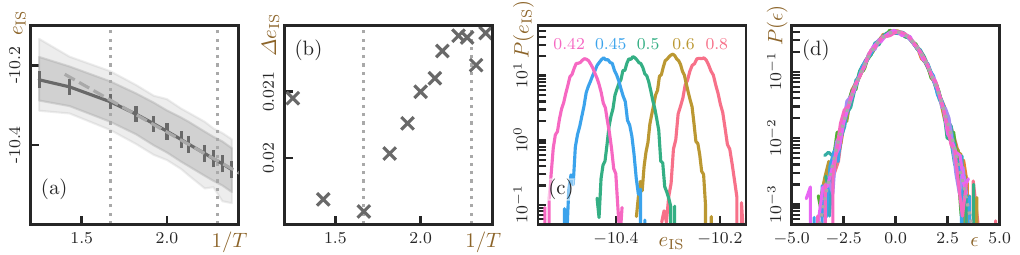}
\end{center}
\caption{
(a) Linear plot of the inherent-structure potential energy per particle, $e_{\rm IS}$, as a function of inverse temperature.
The solid line connects the mean values $\langle e_{\rm IS}\rangle$.
Error bars indicate the standard deviation $\Delta e_{\rm IS}$.
The light shaded area represents the minimum and maximum values, while the dark shaded area indicates the 99th-percentile extreme values.
The dashed line shows the linear relation $\langle e_{\rm IS}\rangle \sim 1/T$ in the low-temperature regime.
(b) Linear plot showing the inverse-temperature dependence of the standard deviation $\Delta e_{\rm IS}$ of $e_{\rm IS}$.
In panels (a) and (b), the dotted vertical lines indicate the locations of $T_{\rm ava}$ and $T_{\rm MCT}$.
(c) Semi-log plot of the probability distribution function $P(e_{\rm IS})$ for $e_{\rm IS}$ at different temperatures.
The temperature corresponding to each curve is indicated above the peak.
(d) Semi-log plot of the probability distribution $P(\epsilon)$ of the rescaled energy $\epsilon$ (defined by Eq.~\ref{eq:epsilon}).
Colors correspond to the same temperatures as in panel (c).
The light-gray dotted line represents the standard normal distribution.
}
\label{fig:E0}
\end{figure*}

\subsection{Energy levels of inherent structures}\label{sec:e_is}
Finally, we characterize the PEL from the perspective of the inherent-structure energy per particle, $e_{\rm IS}$ (Fig.~\ref{fig:PEL}(c)).
The temperature dependence of $e_{\rm IS}$ was first reported in Ref.~\cite{sastrySignaturesDistinctDynamical1998b}, and has since been examined in several subsequent studies~\cite{lanaveRelationLocalDiffusivity2006,dasCrossoverDynamicsKobAndersen2022a}.
In the present work, in addition to the mean value, we measure a variety of statistical quantities, including the standard deviation, the minimum and maximum values, and the 99th-percentile extreme values.
The results are shown in Fig.~\ref{fig:E0}(a).
Although the quantitative values do not coincide because the specific form of the potential-smoothing function is different, the mean value $\langle e_{\rm IS}\rangle$ decreases monotonically with decreasing temperature, in agreement with previous studies~\cite{sastrySignaturesDistinctDynamical1998b,lanaveRelationLocalDiffusivity2006,dasCrossoverDynamicsKobAndersen2022a}.
In particular, consistent with the results of Ref.~\cite{dasCrossoverDynamicsKobAndersen2022a}, a linear dependence on $1/T$ is observed for $T \le T_{\rm ava}\approx 0.6$.
Interestingly, as for several observables reported so far, a qualitative change is observed around $T_{\rm ava}$.
While slight fluctuations are observed in the maximum and minimum values, we stress that the 99th-percentile values smoothly follow the mean value $\langle e_{\rm IS}\rangle$.

Next, Fig.~\ref{fig:E0}(b) shows the standard deviation $\Delta e_{\rm IS}$ of $e_{\rm IS}$ as a function of inverse temperature.
Again, a qualitative change is observed at $T = T_{\rm ava} \approx 0.6$: $\Delta e_{\rm IS}$, which decreases upon cooling at higher temperatures, turns upward below $T_{\rm ava}$.
Furthermore, $\Delta e_{\rm IS}$ levels off in the vicinity of $T_{\rm MCT}$.
The relation between this behavior and avalanche criticality is discussed in Sec.~\ref{sec:picture_ava} in connection with several previous studies.

To examine in more detail whether any qualitative change is present, we further measure the probability distribution functions of $e_{\rm IS}$ at several temperatures, as shown in Fig.~\ref{fig:E0}(c).
The distributions remain approximately Gaussian while shifting toward lower energies upon cooling.
In Fig.~\ref{fig:E0}(d), to make the Gaussianity more evident, we plot the probability distributions of the rescaled energy $\epsilon$, defined using the mean value $\langle e_{\rm IS}\rangle$ and the standard deviation $\Delta e_{\rm IS}$ as
\begin{align}
\epsilon \equiv \frac{e_{\rm IS}-\langle e_{\rm IS}\rangle}{\Delta e_{\rm IS}} .\label{eq:epsilon}
\end{align}
All the distributions closely follow the standard normal distribution indicated by the dotted line, confirming their Gaussian nature.
Thus, from the perspective of the probability distributions, no qualitative change such as the emergence of non-Gaussianity is observed over the entire temperature range considered.
The linear dependence of $\langle e_{\rm IS}\rangle$ on $1/T$ and the Gaussianity of the probability distributions are consistent with the Gaussian landscape model~\cite{sciortinoPotentialEnergyLandscape2005a}.
Note, however, that the standard deviation is predicted to be independent of temperature in the Gaussian model, which is at variance with our results.

\section{Discussions \& overview}\label{sec:dis}
In this section, we discuss the relation between the results reported in this work and various previous studies.
First, in Sec.~\ref{sec:xi}, we first recapitulate the thermal avalanche picture, including the real-space interpretation of the avalanche critical correlation length $\xi$, introduced in Ref.~\cite{tahaeiScalingDescriptionDynamical2023a} on the basis of the T-EPM.
We then explain that, when the results reported in Refs.~\cite{coslovichDynamicThermodynamicCrossover2018a,dasCrossoverDynamicsKobAndersen2022a} are interpreted in the context of avalanche criticality, they suggest a saturation of $\xi$ in the vicinity of the MCT point.
We provide a physical picture for this upper bound of $\xi$ as well.
In Sec.~\ref{sec:ava_PEL}, after summarizing previous studies on the PEL picture of the dynamics of supercooled liquids, we present a PEL-based interpretation of avalanche criticality that is consistent with the results of those previous studies.
In Sec.~\ref{sec:crystal}, we remark on the possible competition between the MCT crossover and crystallization.
Finally, in Sec.~\ref{sec:ubiquity}, we comment on the possibility that the results reported in this work may not be universally applicable to all supercooled-liquid systems.

\begin{figure*}[t]
\begin{center}
\includegraphics[width=\linewidth,angle=0]{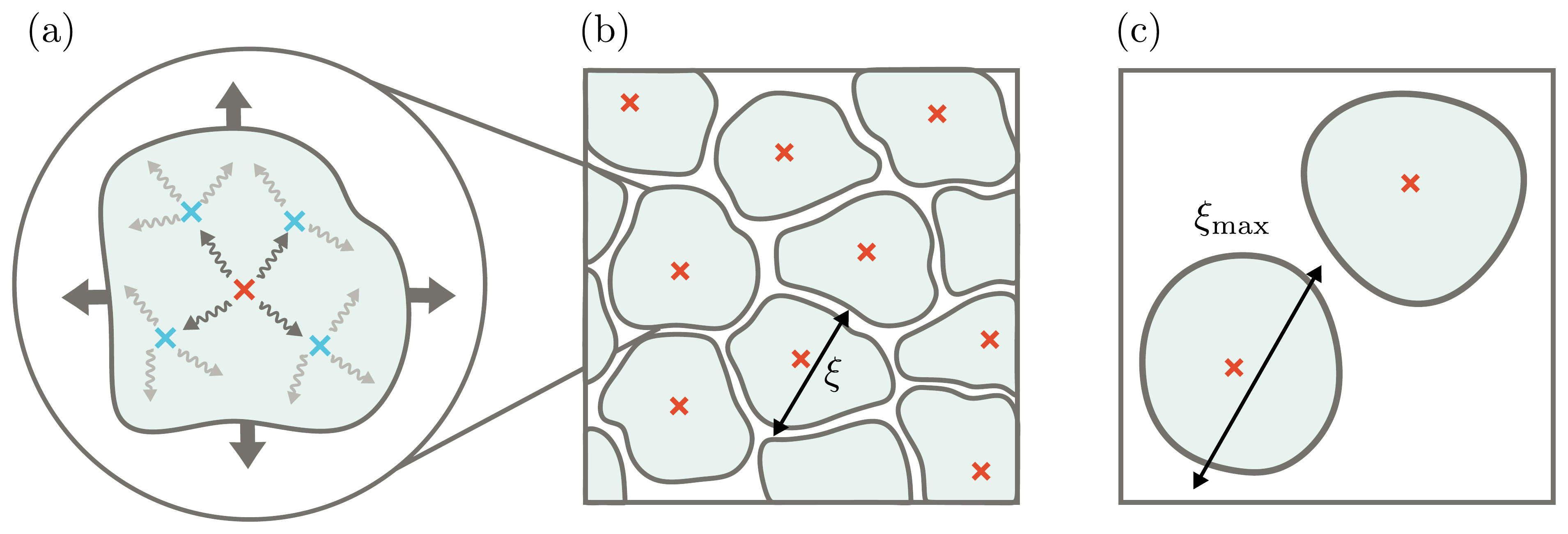}
\end{center}
\caption{
(a) Schematic illustration showing the growth process of an avalanche.
The red cross at the center represents the primary event that triggers the entire avalanching structural relaxation.
Arrows with wavy lines represent the elastic propagation of released energy, and light-blue crosses denote secondary events.
(b) Schematic illustration of a situation at high temperature, where many avalanches occur simultaneously.
This corresponds to a high-density, jammed state of soft avalanche quasiparticles, SAQs (see the main text for the definition).
In this regime, the correlation length is determined by the SAQs density.
(c) Schematic illustration of a situation at low temperature, where only a very small number of avalanches are present.
This corresponds to a low-density, unjammed state of SAQs.
For clarity of visualization, the sketches are shown in two dimensions, although the systems studied in this work are three-dimensional.}
\label{fig:xi_sketch}
\end{figure*}

\subsection{Avalanche critical correlation length}\label{sec:xi}
The physical picture and the mechanism determining the critical correlation length of thermal avalanches have been discussed in Ref.~\cite{tahaeiScalingDescriptionDynamical2023a}.
Since this discussion will play an important role in what follows, we briefly recap it here to make the present article self-contained.

\subsubsection{Thermal avalanches}\label{sec:t_avalanches}
In Ref.~\cite{tahaeiScalingDescriptionDynamical2023a}, a thermal avalanche picture of DH was proposed based on the T-EPM.
We first outline elastoplastic models (EPMs), which provide the basis for the T-EPM.
Next, we explain the thermal avalanche picture proposed in Ref.~\cite{tahaeiScalingDescriptionDynamical2023a}.
Finally, we discuss the consistency between our analysis and the thermal avalanche picture.

EPMs are lattice-based models devised to describe the mechanical response of amorphous solids under shear and they are known to reproduce well the behavior observed in particle-based simulations~\cite{nicolasDeformationFlowAmorphous2018}.
In EPMs, each site is characterized by a stability parameter $x$, defined as the difference between the local yield stress and the current local stress.
Under applied shear, sites for which $x$ reaches zero become unstable, triggering a local rearrangement.
As a result, energy is released through the local structural relaxation.
The released energy propagates through the elastic field and modifies the values of $x$ at surrounding sites depending on their relative positions (Fig.~\ref{fig:xi_sketch}(a)).
When the resulting value of $x$ at a surrounding site becomes zero (or even negative) due to such elastic interactions, a cascade of multiple plastic events can be triggered, and the resulting sequence of rearrangements is referred to as an avalanche.

In Ref.~\cite{ozawaElasticityFacilitationDynamic2023a}, the T-EPM, which extends EPMs to thermal relaxation, was proposed.
In the T-EPM, unlike conventional shear-driven EPMs, each site is probabilistically destabilized by thermal fluctuations according to an Arrhenius-type activation effects associated with the stability parameter $x$.
The energy released by the destabilization of a given site changes the values of $x$ of surrounding sites via the elastic field, as in conventional EPMs; however, it is qualitatively different in that this effect does not necessarily trigger secondary events immediately.
In the equilibrium state of the T-EPM, the probability distribution $P(x)$ of the stability parameter $x$ exhibits a gap near $x=0$ and takes nonzero values only for $x > x_c$.
Here, $x_c$ denotes a positive threshold that depends on temperature.
Due to elastic interactions, the influence of a primary event can cause some sites to satisfy $x < x_c$, thereby significantly accelerating the relaxation times of these secondary sites.
In the thermal avalanche picture proposed in Ref.~\cite{tahaeiScalingDescriptionDynamical2023a}, a sequence of structural relaxations affected by such acceleration effects is regarded as an avalanche.

On the other hand, in the present study, we extract one of the critical exponents characterizing avalanche criticality, $\gamma$, from the number of unstable modes of a saddle configuration, $N^\dagger_{\rm saddle}$.
The fact that structural information about avalanches is encoded in a single configuration may appear to be qualitatively different from the thermal avalanche picture discussed above, in which the sequence of local rearrangements composing an avalanche does not occur simultaneously.
Moreover, since unstable modes are present in our system, the situation may also seem qualitatively different from T-EPM systems, which are characterized by a gapped and totally stable $P(x)$.
However, when entropy effects discussed in Refs.~\cite{hanggiReactionrateTheoryFifty1990,baity-jesiRevisitingConceptActivation2021} are taken into account, the two pictures can be interpreted as consistent.
Although, in the T-EPM, the occurrence of local relaxation events is assumed to be determined purely energetically, in particle-based systems, entropic contributions are clearly non-negligible.
The relevance of entropy effects can be understood, for example, from the fact that even systems at low temperatures, where $\tau_\alpha$ is more than four orders of magnitude larger than the microscopic time scale $\sqrt{\frac{m\sigma^2}{\epsilon}}=1$ (see, e.g., results for $T \le T_{\rm MCT}$), possess a nonzero number of negative modes (Fig.~\ref{fig:2mode}(a) and Fig.~\ref{fig:Ndagger}).
This implies that, even when unstable modes are present, the system does not immediately relax along the directions of these modes.
On the other hand, when relaxation occurs along the direction of any one of these unstable modes, a nonlinear structural rearrangement takes place.
During this process, dynamical facilitation is induced, making rearrangements along the directions of other nearby unstable modes more likely, and the relaxation of the system is then expected to be accelerated.
This scenario is in full agreement with the thermal avalanche picture proposed in Ref.~\cite{tahaeiScalingDescriptionDynamical2023a}.

Despite its simplicity, the T-EPM successfully reproduces the growth of DH in supercooled liquids at low temperatures and provides important insights, in particular into the role of elastic interactions.
In Ref.~\cite{tahaeiScalingDescriptionDynamical2023a}, a scaling argument based on an avalanche criticality picture was proposed for the parameter ($N$ and $T$) dependence of the dynamical susceptibility in the T-EPM, and it was also shown that this picture can account for numerical results with high accuracy.
In the next subsection, we explain the real-space picture of the critical correlation length $\xi$ of thermal avalanches.

\begin{figure*}[t]
\begin{center}
\includegraphics[width=.8\linewidth,angle=0]{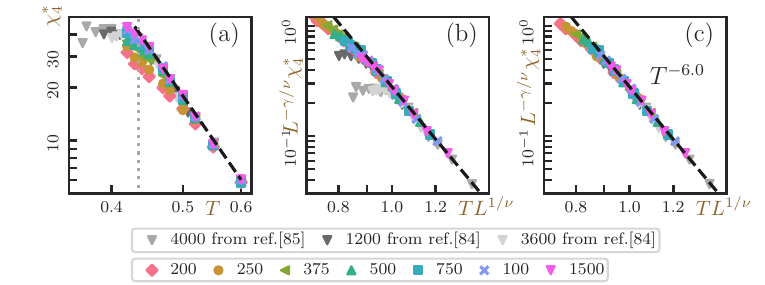}
\end{center}
\caption{
Comparison between data from previous studies~\cite{coslovichDynamicThermodynamicCrossover2018a,dasCrossoverDynamicsKobAndersen2022a} and our results for the peak value of the dynamical susceptibility, $\chi_4^\ast$.
(a) Log-log plot of $\chi_4^\ast$ as a function of temperature.
(b) Finite-size scaling of $\chi_4^\ast$.
(c) Finite-size scaling of $\chi_4^\ast$ including only data points above the MCT transition temperature.
In all panels, only data for $T\le T_{\rm ava}$ are shown.
Dashed lines indicate the power-law behavior $\chi_4^\ast\sim T^{-\gamma}$, and different symbols correspond to different system sizes as indicated in the legend below the panels.
For the data taken from Ref.~\cite{dasCrossoverDynamicsKobAndersen2022a}, the values are rescaled by a factor of $N/N_A=5/4$ to ensure consistency in the definition.
In panel (a), the vertical dashed line marks the position of $T_{\rm MCT}\approx 0.435$.
}
\label{fig:chi4_withprevious}
\end{figure*}

\begin{figure}[t]
\begin{center}
\includegraphics[width=.7\linewidth,angle=0]{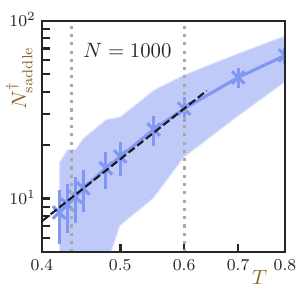}
\end{center}
\caption{
Log-log plot of the number of unstable modes at saddle configurations, $N_{\rm saddle}^\dagger$, as a function of temperature for the system with $N = 1000$. Symbols represent the mean values, error bars indicate the standard deviations, and the shaded region shows the range between the maximum and minimum values. The light-gray vertical dotted lines indicate the locations of $T_{\rm ava} \approx 0.6$ and $T_{\rm MCT} \approx 0.435$. The black dashed line shows the power-law behavior in the critical regime ($T_{\rm MCT} \le T \le T_{\rm ava}$).
}
\label{fig:Ndagger_loglog}
\end{figure}

\subsubsection{Saturation of the correlation length}\label{sec:xi_max}
At sufficiently high temperatures, multiple independent avalanches can occur simultaneously~\cite{whiteDrivingRateEffects2003a}.
In such cases, individual avalanches interfere with each other before they can fully develop, leading to a truncation of the correlation length (Fig.~\ref{fig:xi_sketch}(b)).
The typical spatial extent of an avalanche determined by this mechanism provides a real-space picture of the correlation length $\xi$ at a given temperature.

In Refs.~\cite{coslovichDynamicThermodynamicCrossover2018a,dasCrossoverDynamicsKobAndersen2022a}, the relaxation dynamics of supercooled liquids were measured down to temperatures significantly lower than the MCT transition point.
As shown in Fig.~\ref{fig:chi4_withprevious}(a), these studies reported that the system-size dependence of $\chi_4^\ast$ disappears for $N \ge 1200$.
In terms of temperature, this disappearance of system-size dependence corresponds to a saturation of $\chi_4^\ast$ (and thus of the correlation length $\xi$) in the vicinity of $T_{\rm MCT} \approx 0.435$.
As a consequence of this saturation, the finite-size scaling breaks down at temperatures below the MCT point, as shown in Fig.~\ref{fig:chi4_withprevious}(b).
For reference, Fig.~\ref{fig:chi4_withprevious}(c) shows the scaling results obtained by using only data at $T \ge T_{\rm MCT} \approx 0.435$, including the data for $N \ge 1200$ taken from Refs.~\cite{coslovichDynamicThermodynamicCrossover2018a,dasCrossoverDynamicsKobAndersen2022a}.
In this temperature range, the results for all system sizes collapse with high precision under finite-size scaling.

Two possible origins can be considered for the saturation of $\xi$ around the MCT point. 
The first is that the number of unstable modes, $N_{\rm saddle}^\dagger$, ceases to decrease near the MCT point (recall that $N_{\rm saddle}^\dagger$ serves as a measure of the avalanche density, and that the avalanche density determines $\xi$). 
The second possibility is that there exists an upper bound on the dynamical correlation length that is independent of $N_{\rm saddle}^\dagger$.
The first possibility, concerning $N_{\rm saddle}^\dagger$, can be directly tested using the simulation results.
Figure~\ref{fig:Ndagger_loglog} presents a log-log plot of the same $N_{\rm saddle}^\dagger$ data for the $N=1000$ system shown in Fig.~\ref{fig:Ndagger}(c).
From this result, $N_{\rm saddle}^\dagger$ continues to decrease while following essentially the same scaling even below the MCT point.
This indicates that the saturation of $\xi$ originates from the presence of an intrinsic upper bound that is independent of $N_{\rm saddle}^\dagger$.

We now consider a physical picture for the existence of an intrinsic upper bound $\xi_{\rm max}$.
Even at sufficiently low temperatures, avalanches occur simultaneously at low density if the system size is sufficiently large.
Because these avalanches are well separated spatially, they can grow to large sizes without mutual interference.
The existence of $\xi_{\rm max}$, however, implies that the growth of an avalanche can spontaneously cease even in the absence of interference between avalanches.
This puzzling behavior can be understood intuitively by regarding the spatial extent of an avalanche as a soft quasiparticle, which we refer to as a soft avalanche quasiparticle (SAQ).
At sufficiently high temperatures, SAQs exist at high density and are jammed.
In such a situation, compression between SAQs effectively reduces their sizes, resulting in a correlation length smaller than $\xi_{\rm max}$.
By contrast, the approach of $\xi$ to $\xi_{\rm max}$ can be interpreted as an unjamming-like phenomenon in which the interaction network among SAQs becomes disconnected\footnote{We emphasize that, since the translational degrees of freedom of SAQs are not expected to be coupled to their interactions, the behavior discussed here should not follow jamming criticality. Rather, we use the term unjamming in a more conceptual sense, referring to the collective loss of the interaction (contact) network.}.
Under such unjammed conditions, each SAQ acquires an intrinsic size $\xi_{\rm max}$, as illustrated in Fig.~\ref{fig:xi_sketch}(c).

From the perspective of the unjamming of SAQs, the localization of unstable modes discussed in Sec.~\ref{sec:deloc} can also be understood.
As recapped in Sec.~\ref{sec:chi4}, our scaling argument is based on the hypothesis that each unstable mode originally corresponds to a thermally excited QLV (i.e., an STZ).
The success of finite-size scaling with the critical exponents predicted by this hypothesis (Figs.~\ref{fig:chi4star} and~\ref{fig:chi4_withprevious}) further supports its validity.
On the other hand, modes with closely spaced eigenvalues can hybridize~\cite{rainonePinchingGlassReveals2020,lernerEnumeratingLowfrequencyNonphononic2024}.
Accordingly, while an excited QLV itself has a spatially localized structure, in a jammed state of SAQs the hybridization of multiple modes can lead to the formation of an extended single mode.
Below the MCT point, however, SAQs become unjammed and only localized avalanche structures remain in the system; consequently, even if all unstable modes were to hybridize, they would still remain localized.
Within this picture, the saturation of the correlation length and the localization of unstable modes near the MCT point discussed in Sec.~\ref{sec:deloc} are expected to occur at the same temperature, as both arise from the unjamming of SAQs.
Moreover, if extended modes originate from the hybridization of multiple unstable modes, it is natural to expect that such modes exhibit a fractal structure similar to that of avalanches.
For this reason, the working hypothesis adopted in Sec.~\ref{sec:fractal_me}, namely that the fractal dimension of the unstable modes equals that of avalanches, $d_f$, appears justified.

We emphasize that the saturation of the dynamical susceptibility (and, accordingly, the correlation length) discussed in this subsection signifies a breakdown of avalanche criticality in the vicinity of the MCT point.
In other words, in the deeply supercooled regime below the MCT point, the dominant factor governing slow dynamics is expected to be distinct from thermal avalanches.
Therefore, our results do not immediately indicate that the glass transition occurs at zero temperature, but rather imply the emergence of a new dominant mechanism of slow dynamics in the low-temperature regime.

\subsection{Potential energy landscape picture of avalanche criticality}\label{sec:ava_PEL}
In this section, we first summarize the understanding of the PEL developed in previous studies in Sec.~\ref{sec:PEL_previous}.
We then present an interpretation of avalanche criticality based on the PEL picture in Sec.~\ref{sec:picture_ava}.
In addition, we mention an important open question related to the PEL picture in Sec.~\ref{sec:open_PEL}.

\subsubsection{Overview of previous studies on the PEL}
\label{sec:PEL_previous}
The PEL picture of supercooled liquids has been extensively discussed in many previous studies.
In this subsection, we review several of these works that are particularly relevant to understanding avalanche criticality.

\emph{Metabasin jumps and entropy effects---.}
To directly investigate how the dynamics of supercooled liquids under thermal fluctuations is influenced by the PEL, a method known as inherent dynamics has been employed, in which inherent structures associated with instantaneous configurations are tracked at regular intervals during a standard molecular simulations~\cite{lemaitreStructuralRelaxationScaleFree2014,takahaAvalancheCriticalityEmerges2025}.
Multiple studies using inherent dynamics have demonstrated that structural relaxation in supercooled liquids proceeds predominantly through transitions between metabasins~\cite{doliwaEnergyBarriersActivated2003,appignanesiDemocraticParticleMotion2006}.
As mentioned in Sec.~\ref{sec:t_avalanches}, it has also been argued that the dynamics is not determined solely by energy barriers, but that entropic effects arising from the scarcity of energetically favorable pathways also play an important role~\cite{baity-jesiRevisitingConceptActivation2021}.

\emph{MCT (Mode-coupling theory)---.} 
MCT is a theoretical framework that predicts a dynamical transition~\cite{janssenModeCouplingTheoryGlass2018}.
At the transition point predicted by MCT, the so-called MCT point, a nonergodic transition occurs, in which transitions between different basins become forbidden.
This behavior can be viewed as the emergence of certain energy-landscape structures, such as metabasins.
However, this nonergodic transition arises from the divergence of energy barriers in infinite-dimensional models.
In finite-dimensional systems, only finite energy barriers can form; as a result, a complete nonergodic transition does not occur even at temperatures below the MCT point, and hopping processes can still take place.
Although MCT and the avalanche picture are similar in the sense that both involve dynamical transitions, they focus on different aspects: MCT describes a transition in which hopping is suppressed, whereas the avalanche picture is based on dynamics initiated by hopping events.

In early studies on finite-dimensional properties related to MCT, the eigenvalue spectra of normal modes at saddle configurations were investigated, and it was reported that unstable (negative) modes disappear at the MCT point~\cite{angelaniSaddlesEnergyLandscape2000a,broderixEnergyLandscapeLennardJones2000b}.
This result is in line with the predictions of mean-field theory~\cite{cavagnaRoleSaddlesMeanfield2001} and the expectation that, below the MCT point, the dynamics of finite-dimensional supercooled liquids is dominated by hopping processes.
However, as discussed above in Sec.~\ref{sec:deloc}, this saddle-disappearance scenario at the MCT point was ruled out by Ref.~\cite{coslovichLocalizationTransitionUnderlies2019b}, which demonstrated that unstable modes do not disappear at the MCT point but instead become fully localized.

\emph{Replica theory---.}
Replica theory provides a thermodynamic perspective on metabasins~\cite{parisiInfiniteNumberOrder1979,monassonStructuralGlassTransition1995,franzRecipesMetastableStates1995}.
Since this theory is based on thermodynamic arguments, it is fundamentally different from MCT, which is a dynamical theory.
However, in simple mean-field models such as the $p$-spin spherical model, where both thermodynamic and dynamical analyses can be performed, the correspondence between MCT and replica theory can be made explicit.
For example, it can be shown that the MCT point lies at a higher temperature than the transition associated with replica-symmetry breaking (corresponding to the Kauzmann temperature)~\cite{castellaniSpinglassTheoryPedestrians2005}.
Note that, since replica theory describes equilibrium states, its predictions pertain to the free energy landscape rather than the PEL.

Within the framework of replica theory, extensive studies have also been carried out on structural glass systems based on replica liquid theory~\cite{parisiTheorySimpleGlasses2020,parisiMeanfieldTheoryHard2010,coluzziThermodynamicsBinaryMixture1999,mezardThermodynamicsGlassesFirst1999a,franzEffectivePotentialGlassy1998}.
In particular, following the prediction of the Gardner transition from a stable glass (the one-step replica-symmetry-breaking, 1RSB, phase) to a marginally stable glass (the full replica-symmetry-breaking, full-RSB, phase)~\cite{charbonneauFractalFreeEnergy2014,biroliLiuNagelPhaseDiagrams2018}, intensive efforts have been devoted to its verification through numerical simulations in finite-dimensional systems~\cite{berthierGrowingTimescalesLengthscales2016,scallietAbsenceMarginalStability2017a,hicksGardnerTransitionPhysical2018,seoaneLowtemperatureAnomaliesVapor2018,scallietRejuvenationMemoryEffects2019a,scallietMarginallyStablePhases2019} as well as through experiments~\cite{koolGardnerlikeCrossoverVariable2022,seguinExperimentalEvidenceGardner2016,hammondExperimentalObservationMarginal2020,xiaoProbingGardnerPhysics2022}.
The Gardner transition refers to the fragmentation of a metabasin in the free energy landscape into a hierarchy of subbasins, whereby nearly identical configurations become distinguished as distinct states.
Recent studies have reported that, in finite-dimensional systems, rather than a sharp phase transition accompanied by a divergence of a correlation length, full-RSB-like hierarchical organization is observed as a sharp crossover characterized by peaks in susceptibilities~\cite{liaoDynamicGardnerCrossover2023,liaoHierarchicalLandscapeHard2019a}.

While conventional replica theory is formulated for the free energy, Ref.~\cite{franzMeanfieldAvalanchesJammed2017} considered a PEL with a hierarchical structure expected for a full-RSB state.
In that work, it was theoretically shown that plastic deformation induced by applying an infinitesimal shear to a static configuration on such a full-RSB-like hierarchical PEL exhibits avalanche criticality.
The connection between full-RSB-type marginal stability and shear-induced avalanche criticality has also been discussed by means of numerical simulations of particle systems~\cite{shangElasticAvalanchesReveal2020,oyamaShearinducedCriticalityGlasses2023a}.

\emph{Temperature dependence of the inherent-structure energy---.}
As mentioned above in Sec.~\ref{sec:e_is}, the inherent-structure energy per particle, $e_{\rm IS}$, has been measured in previous studies~\cite{sastrySignaturesDistinctDynamical1998b,dasCrossoverDynamicsKobAndersen2022a,lanaveRelationLocalDiffusivity2006}.
Above the onset temperature ($T_{\rm onset} \approx 1.0$), $e_{\rm IS}$ is essentially independent of temperature, whereas below this temperature it decreases monotonically upon cooling. 
For the KAM, Ref.~\cite{dasCrossoverDynamicsKobAndersen2022a} performed measurements down to temperatures as low as $T = 0.365$.
That work showed that, for $T < 0.7$, the average inherent-structure energy $\langle e_{\rm IS}\rangle$ decreases linearly with $1/T$ down to $T = 0.365$.

As a theoretical approach to the temperature dependence of $e_{\rm IS}$, analyses based on the mean-field $p$-spin spherical model have been developed~\cite{cugliandoloAnalyticalSolutionOffequilibrium1993}.
Within the pure $p$-spin spherical model, it is predicted that the inherent-structure energy remains constant for parent temperatures above the MCT point.
This prediction is in contradiction with numerical simulation results for the KAM, and this discrepancy remained unresolved for many years.
Only recently, analyses based on mixed $p$-spin spherical models have shown that $e_{\rm IS}$ starts to decrease from an onset temperature located above the MCT point, in qualitative agreement with numerical results for the KAM~\cite{folenaRethinkingMeanFieldGlassy2020}.
In contrast to this progress in understanding the behavior of the average inherent-structure energy $e_{\rm IS}$, the temperature dependence of $\Delta e_{\rm IS}$ (reported in Fig.~\ref{fig:E0}(b)) remains theoretically unexplained, to the best of our knowledge.

\emph{Activation energy---.} 
Another important measure for quantitative characterization of the PEL is the activation energy.
The temperature dependence of the activation energy in the KAM has been investigated in several studies.
Confusingly, the reported behavior at low temperatures is not consistent across the literature even if we restrict ourselves to the KAM: some studies report that the activation energy increases upon cooling~\cite{coslovichDynamicThermodynamicCrossover2018a}, whereas others claim a sharp decrease~\cite{baity-jesiRevisitingConceptActivation2021}.
More recently, several refined methods for measuring activation energies have been proposed and applied to swap models~\cite{ninarelloModelsAlgorithmsNext2017a}, enabling precise measurements of the activation energy~\cite{picaciamarraLocalVsCooperative2024,jiRoleExcitationsSupercooled2025}.
According to these studies, at least for swap models, the activation energy increases at low temperatures.

\subsubsection{Avalanche criticality}\label{sec:picture_ava}
In this subsection, we present a PEL picture of avalanche criticality that is consistent with the established understanding summarized in Sec.\ref{sec:PEL_previous}.
In our interpretation, an avalanche corresponds to a sequence of correlated structural relaxations initiated by hopping between metabasins~\cite{doliwaEnergyBarriersActivated2003,appignanesiDemocraticParticleMotion2006}.
This process can be viewed as a descent along the surface of a metabasin in the PEL.
We further propose that a hierarchical organization (referred to as subbasins hereafter) exists within a metabasin, and that a thermal avalanche corresponds to the relaxation process by which the system descends through the metabasin while being temporarily trapped in subbasins.
Although an infinite hierarchy as in full-RSB may not strictly exist in finite-dimensional and finite-size systems, the presence of a certain degree of hierarchical organization is expected~\cite{liaoDynamicGardnerCrossover2023,liaoHierarchicalLandscapeHard2019a}.
As we explain below, this picture is consistent with known results on avalanche criticality observed in shear-induced plastic events.

As mentioned in Sec.~\ref{sec:PEL_previous}, Ref.~\cite{franzMeanfieldAvalanchesJammed2017} considered the response of systems characterized by a PEL with full-RSB-type hierarchical structure to an external field.
Reflecting the infinite hierarchy of the PEL, the system undergoes rearrangements associated with transitions between subbasins even under an infinitesimal applied strain.
It was theoretically shown that the probability distribution of the energy changes associated with such rearrangements exhibits avalanche criticality.
By contrast, it is known that numerical simulations performed in finite-dimensional systems under similar setups do not display such criticality~\cite{karmakarStatisticalPhysicsYielding2010a,oyamaUnifiedViewAvalanche2021,oyamaShearinducedCriticalityGlasses2023a}.
For example, in Ref.~\cite{oyamaUnifiedViewAvalanche2021}, avalanche size distributions measured under athermal quasistatic shear~\cite{maedaComputerSimulationDeformation1978}~\footnote{An athermal quasistatic shearing protocol means that shear is applied in a quasistatic manner in the absence of thermal fluctuations. It is typically implemented by repeatedly applying small strain increments, each followed by energy minimization.} were compared between two situations: those obtained by considering only the first plastic event observed in as-quenched samples (first-event ensemble), and those observed in the steady state after macroscopic yielding.
The former first-event ensemble can be regarded as corresponding to the setup considered in Ref.~\cite{franzMeanfieldAvalanchesJammed2017}.
However, while the steady-state distributions clearly exhibit avalanche criticality, no system-size dependence was observed for the first-event ensemble.
At the same time, even in the first-event ensemble, the avalanche size distribution displays a power-law-like behavior up to a system-size-independent cutoff size.
The existence of this cutoff size is consistent with the upper bound of the correlation length discussed in Sec.~\ref{sec:xi_max}.
Taking into account the arguments presented in Ref.~\cite{parisiShearBandsManifestation2017}, as discussed below, we suggest that this cutoff size corresponds to the typical size of metabasins.

Using numerical simulations of the two-dimensional KAM, Ref.~\cite{parisiShearBandsManifestation2017} provided more direct evidence for the existence of subbasin-like structures in the PEL, as well as for the importance of jumps between metabasins in structural relaxation.
In Ref.~\cite{parisiShearBandsManifestation2017}, the authors generated different replicas by applying small perturbations to the same initial configuration.
They then investigated how the probability distribution of the structural overlap between these replicas evolves under the application of athermal quasistatic shear strain.
While the overlap distribution gradually shifts toward smaller values with increasing strain, a first-order-like transition has been reported at the yielding strain.
Below the yielding point, large overlap values are observed, indicating that only minor structural changes occur and thus suggesting the presence of subbasins.
Moreover, the large structural change characterized by small overlap values, which emerges in a first-order-like manner, can be interpreted as a jump between metabasins at the yielding strain.

The temperature dependence of $\Delta e_{\rm IS}$, the fluctuations of the inherent structure energy (Fig.~\ref{fig:E0}b), also supports the existence of subbasins.
Within our picture, $\Delta e_{\rm IS}$ is interpreted as reflecting differences in energy among the hierarchical subbasin structures encountered during the descent
on the PEL surface.
At lower temperatures, where $\xi$ becomes larger, the descent process on the PEL surface can access energies associated with various hierarchical levels.
In this view, it is natural that $\Delta e_{\rm IS}$ increases upon cooling.
The existence of an upper bound of $\Delta e_{\rm IS}$ around $T_{\rm MCT}$ is also consistent with the behavior of $\xi$, suggesting that the system has reached the
bottom of the metabasin.
We also note that $\Delta e_{\rm IS}$ within the critical regime increases only by about 10\% (Fig.~\ref{fig:E0}b).
This likely reflects that the energy differences between subbasins are small compared to the energy fluctuations of the metabasin itself in this temperature range.

We conclude this subsection with a PEL picture for the situation where multiple avalanches occur simultaneously.
When multiple avalanches are present, distinct descent directions associated with each avalanche interfere with one another in the high-dimensional space on which the PEL is defined, causing the system to be diverted into a different direction before reaching the minimum of each basin.
As shown in Sec.~\ref{sec:xi}, in real space this corresponds to the mutual interference of nearby avalanches, which suppresses the growth of each avalanche.
As the temperature decreases and the density of avalanches becomes smaller, avalanches can grow without interference.
In other words, the system can descend to lower energies on the PEL.
Finally, when the density of avalanches becomes sufficiently small, avalanches are expected to be able to descend all the way to the minima within a metabasin.
This corresponds to the saturation of the growth of the correlation length (Fig.~\ref{fig:xi_sketch}(c)), and the maximum value of the correlation length can be interpreted as reflecting the spatial extent of the basin of attraction of a metabasin.
Note that even below the MCT point, at which the correlation length reaches its maximum value, multiple avalanches may still occur simultaneously: if the eigenvectors forming each avalanche in phase space reside in mutually orthogonal subspaces, no interference occurs in real space.
Based on our picture of the correlation length, it is therefore expected that below the MCT transition point the volume of the basin of attraction in phase space remains of comparable magnitude.
We emphasize that this behavior is in contrast to the energy levels of inherent structures, which continue to decrease with decreasing temperature even below the MCT point (see Sec.~\ref{sec:e_is}).

To summarize the discussions presented in this section so far, our numerical calculations confirm the validity of the zero-temperature avalanche criticality picture, suggesting that inter-metabasin jumps on the PEL constitute the microscopic physical origin governing glassy slow dynamics.
As we explained, this scenario is consistent with several characteristics of the PEL confirmed in our own numerical calculations, such as the vibrational density of states of inherent structures (Sec.~\ref{sec:vdos}) and the localization properties of unstable modes associated with saddle configurations (Sec.~\ref{sec:deloc}), as well as with many other PEL features reported in previous studies.
On the other hand, as discussed in Sec.~\ref{sec:xi_max}, the critical correlation length $\xi$ ceases to develop below the MCT point and therefore does not appear to govern the dynamics in such deeply supercooled regime.
In other words, our results do not imply that the glass transition occurs purely at zero temperature.

\subsubsection{Activation energy: an open question}\label{sec:open_PEL}
Here, we discuss the temperature dependence of the activation energy $\Delta E$.
As discussed in Sec.~\ref{sec:PEL_previous}, studies on the KAM report conflicting results: one measurement indicates that the activation energy increases upon cooling~\cite{coslovichDynamicThermodynamicCrossover2018a}, whereas another reports a decrease~\cite{baity-jesiRevisitingConceptActivation2021}.
On the one hand, the measurement protocol employed in Ref.~\cite{coslovichDynamicThermodynamicCrossover2018a} was later pointed out to have shortcomings in Ref.~\cite{picaciamarraLocalVsCooperative2024}.
On the other hand, Ref.~\cite{baity-jesiRevisitingConceptActivation2021} investigated systems with a very small system size of $N=65$, raising concerns about finite-size effects.
Thus, neither of these results can yet be regarded as providing a definitive conclusion.

We therefore estimated the activation energy of the KAM using the reheating protocol proposed recently in Ref.~\cite{picaciamarraLocalVsCooperative2024}.
In the reheating method, the $\alpha$ relaxation time is assumed to follow an Arrhenius law with a temperature-dependent activation energy $\Delta E$. 
The value of $\Delta E$ is then determined from the macroscopic relaxation behavior of the system.
However, as shown in Appendix~\ref{ap:deltaE}, the results obtained in the present measurements indicate that the microscopic time scale, which appears as the prefactor in the Arrhenius law, exhibits a temperature dependence that is physically unnatural.
This unexpected temperature dependence of the microscopic time scale suggests that a single Arrhenius law may not provide an adequate description of the macroscopic relaxation in the present system. 
This behavior may reflect entropy effects discussed above in Sec.~\ref{sec:t_avalanches} and \ref{sec:PEL_previous}, as well as the spatiotemporal heterogeneity of multiple relaxation processes.
It is currently unclear to what extent the values of the activation energy obtained from our analysis can be considered reliable.
For this reason, the details of the procedure and the results of the reheating method are presented in Appendix~\ref{ap:deltaE}, not in the main text.
In addition to the reheating protocol, Refs.~\cite{picaciamarraLocalVsCooperative2024,jiRoleExcitationsSupercooled2025} proposed measurement methods based on single hopping events, such as Systematic Excitation ExtRaction (SEER) and Athermal Systematic Excitation ExtRaction (ASEER).
Estimating the activation energy using these approaches constitutes important future work.

\subsection{Influence of crystallization}\label{sec:crystal}

As explained in Appendix~\ref{ap:crystal}, the discussion in the present study is restricted to a temperature range in which crystallization does not pose a problem.
On the other hand, it is also true that, in the KAM, crystallization starts to become significant at temperatures below the MCT point~\cite{coslovichDynamicThermodynamicCrossover2018a,ingebrigtsenCrystallizationInstabilityGlassForming2019a,dasCrossoverDynamicsKobAndersen2022a}.
Consistently, in our measurements, a small number of crystallized samples are indeed detected for that temperature range (see Appendix~\ref{ap:detection}).
The fact that the saturation of the correlation length and the onset of crystallization approximately coincide at the MCT point may be purely coincidental; however, based on the current state of knowledge, this issue remains unclear.

Extending the present analysis to systems with strongly suppressed crystallization remains an important direction for future work.
For example, the variant of the KAM with a particle-type number ratio of $2{:}1$ proposed in Ref.~\cite{ortliebProbingExcitationsCooperatively2023} is essentially the same system as the KAM employed in the present study, while eliminating the issue of crystallization, and is thus well suited as a model system for validation.

Furthermore, it is also important to investigate the swap system~\cite{ninarelloModelsAlgorithmsNext2017a}, which is qualitatively different from the KAM in the sense that it lacks attractive interactions.
As will be discussed in the next subsection, qualitatively different behavior has already been observed in the low-frequency limit of the vibrational density of states.

\subsection{Ubiquity of avalanche-induced dynamical heterogeneity}\label{sec:ubiquity}
Finally, in this subsection, we discuss the ubiquity of the findings of the present study based on results reported in previous works.
As reported in Sec.~\ref{sec:critical_regime}, the upper bound of the temperature range over which avalanche criticality holds was identified using $A_{\rm S}$ extracted from
$D_0^{\rm S}(\omega)=A_{\rm S}\omega^{6.5}$.
In the KAM, avalanche criticality was found to hold below a temperature $T_{\rm ava}$ at which the value of $A_{\rm S}$ starts to decrease.
Furthermore, as explained in Sec.~\ref{sec:xi_max}, the correlation length associated with this avalanche criticality is found to saturate below $T_{\rm MCT}$.

On the other hand, in Ref.~\cite{wangLowfrequencyVibrationalModes2019a}, the temperature dependence of the quartic law
$D_0(\omega)=A_4\omega^4$
in the swap system~\cite{ninarelloModelsAlgorithmsNext2017a} was investigated, and it was shown that $A_4$ remains unchanged in the high-temperature regime above the MCT point, whereas it starts to decrease upon cooling below the MCT transition point (we note that Ref.~\cite{xuLowfrequencyVibrationalDensity2024b} confirmed that $A_4$ and $A_{\rm S}$ exhibit qualitatively consistent temperature dependence).
This implies that, in the swap system, the temperature $T_{\rm ava}$ at which stability starts to change coincides with $T_{\rm MCT}$ at which $\chi_4^\ast$ saturates~\cite{shiraishiCharacterizingSlowDynamics2024}.
Taken together with our findings for the KAM, these results suggest that, in the swap system, there may be no temperature regime in which DH follows avalanche criticality.
In other words, in contrast to the broad ubiquity of DH itself, its physical origin may depend sensitively on the details of the system. 
Therefore, it is important and interesting to apply the same analyses as in the present study to the swap system.
We note that, for the swap system, the relevance of elastoplastic dynamical facilitation below the MCT point has been reported~\cite{chackoElastoplasticityMediatesDynamical2021a,scallietThirtyMillisecondsLife2022}.
We also note that DH estimated under meta-dynamics such as swap Monte Carlo is known to be qualitatively different from that observed under the physical dynamic rule~\cite{shiraishiCharacterizingSlowDynamics2024}.

\section{Conclusion}\label{sec:conclusion}
In this article, we investigated the slow dynamics of the three-dimensional KAM, a prototypical model of supercooled liquids, by means of molecular dynamics simulations. 
We first demonstrated that the temperature and system-size dependence of two representative indicators of slow dynamics, the overlap function and the dynamical susceptibility, can be explained in a unified manner within the zero-temperature avalanche criticality picture.
The validity of this criticality picture suggests that the dynamical properties of the KAM are governed by the PEL. 
We therefore quantitatively characterized the PEL from three distinct perspectives: the vibrational density of states of inherent structures, the spatial localization of unstable modes in saddle configurations, and the energy levels of inherent structures.
Based on the insights obtained from these analyses of the PEL, we further propose a PEL-based interpretation of avalanche criticality that is consistent with various previous studies.
Within our proposed picture, not only the temperature and system-size dependence of $Q(t)$ and $\chi_4(t)$, but also previously unexplained phenomena observed near the MCT point, such as the saturation of the dynamical susceptibility~\cite{coslovichDynamicThermodynamicCrossover2018a,dasCrossoverDynamicsKobAndersen2022a} and the localization of unstable modes in saddle configurations~\cite{coslovichLocalizationTransitionUnderlies2019b}, are naturally unified.
Importantly, the former, namely the saturation of the dynamical susceptibility, implies that the criticality becomes irrelevant in the deeply supercooled regime below $T_{\rm MCT}$.
Therefore, our zero-temperature criticality scenario does not immediately indicate that the glass transition takes place at zero temperature.
Rather, the saturation of the dynamical susceptibility alongside the continued growth of the relaxation time suggests the emergence of a distinct factor governing the dynamics in the deeply supercooled regime.
Identifying such a mechanism will be an important direction for future work toward understanding the nature of the glass transition.

Based on previous studies, it is suggested that our proposed picture may not hold universally for other systems~\cite{wangLowfrequencyVibrationalModes2019a, shiraishiCharacterizingSlowDynamics2024}. 
Therefore, performing similar analyses in different systems, such as the swap system~\cite{ninarelloModelsAlgorithmsNext2017a} and modified KAM models~\cite{ortliebProbingExcitationsCooperatively2023}, constitutes an important direction for future work. 
Furthermore, regarding the temperature dependence of the activation energy, one of the key characteristics of the PEL, no firm conclusions have yet been established for the KAM. 
A reliable determination of the activation energy thus remains an important open problem.
Given the discussions in Refs.~\cite{wyartDoesGrowingStatic2017a,picaciamarraLocalVsCooperative2024,jiRoleExcitationsSupercooled2025}, changes in the local activation energy may constitute the dominant mechanism governing the dynamics in the deeply supercooled regime.

Our analyses have revealed a close relationship between slow glassy dynamics and the fraction of unstable modes at saddle-point configurations.
Given that the normal modes are determined solely by the interaction potential and particle configurations, identifying a structural order parameter that quantifies the local contribution to unstable modes~\cite{liuQuasilocalizedVibrationsArise2025} is expected to provide insight into the structural origin of glassy dynamics~\cite{tanakaCriticallikeBehaviourGlassforming2010,kawasakiCorrelationDynamicHeterogeneity2007,kawasakiStructuralOriginDynamic2010,tongRevealingHiddenStructural2018,tongStructuralOrderGenuine2019,oyamaWhatDeepNeural2023,liuClassificationSolidLiquid2024}.

\begin{acknowledgments}
The authors thank Yuki Takaha, Harukuni Ikeda, Hideyuki Mizuno, Atsushi Ikeda, Hajime Yoshino, and Kunimasa Miyazaki for fruitful discussions.
This work was financially supported by JSPS KAKENHI Grant Numbers JP24H02203, JP24H00192, JP25K00968.
In this research work, we used the “mdx: a platform for building data-empowered society”~\cite{suzumuraMdxCloudPlatform2022}.
\end{acknowledgments}

\appendix

\begin{figure}[t]
\begin{center}
\includegraphics[width=\linewidth,angle=0]{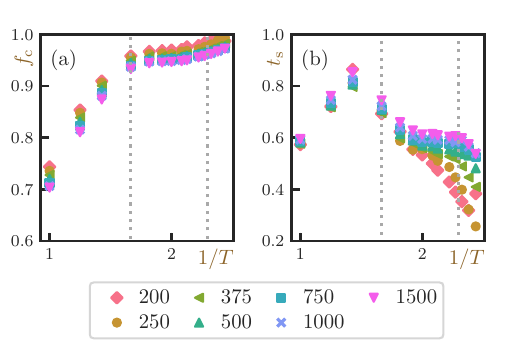}
\end{center}
\caption{
Parameter reflecting the fast mode contribution, obtained from fitting $Q(t)$ using the two-mode relaxation model (Eq.~\ref{eq:2mode}).
(a) Linear plot of the fraction of the slow mode, $f_{\rm C}$, as a function of the inverse temperature.
(b) Linear plot of the fast mode relaxation time, $t_{\rm S}$, as a function of the inverse temperature.
In both panels, the vertical dashed lines indicate the positions of $T_{\rm ava} \approx 0.6$ and $T_{\rm MCT} \approx 0.435$, and the different symbols correspond to different system sizes, as indicated in the legend below the panels.
}
\label{fig:2mode_fast}
\end{figure}

\section{Fast mode parameters}\label{ap:fast}
In this Appendix, we present measurement results of two of the four parameters of the two-mode relaxation model introduced in Sec.~\ref{sec:Q} which were not discussed in the main text: the fraction of the slow mode $f_{\rm C}$ and the characteristic timescale of the fast mode $t_{\rm S}$.

Figure~\ref{fig:2mode_fast}(a) shows $f_{\rm C}$ as a function of $1/T$.
In terms of temperature dependence, $f_{\rm C}$ increases monotonically upon cooling for $T > T_{\rm ava}$, while it exhibits a plateau at temperatures below $T_{\rm ava}$.
It then starts to increase again at temperatures slightly above $T_{\rm MCT}$.
These trends are very similar to those of $\beta_{\rm KWW}$ reported in Fig.~\ref{fig:2mode}(b) of the main text. 
On the other hand, the system-size dependence appears to be largely independent of temperature, with the results for each system size forming approximately parallel shifts, in contrast to the case of $\beta_{\rm KWW}$. 
For $N \ge 1000$, the system-size dependence seems negligible.
The temperature dependence appears to reflect the increase of the Debye-Waller factor at low temperatures~\cite{lewisMoleculardynamicsStudySupercooled1994,scopignoFragilityLiquidEmbedded2003}.
Regarding the $N$ dependence, in smaller systems the number of low-frequency modes is reduced, and the excitation of STs and elastic waves is suppressed. This may explain why $f_{\rm C}$ is higher in smaller systems (i.e., the contribution from the fast mode is smaller).
Although it is intriguing that $f_{\rm C}$ exhibits a plateau roughly in the critical regime $T_{\rm MCT} \le T \le T_{\rm ava}$, a clear understanding of this behavior has not yet been established.

Figure~\ref{fig:2mode_fast}(b) shows the characteristic timescale of the fast mode, $t_{\rm S}$. 
Similar to $f_{\rm C}$, $t_{\rm S}$ exhibits temperature and system-size dependences that are similar to those of $\beta_{\rm KWW}$.
However, slight differences are observed: the formation of the plateau in $t_{\rm S}$ is shifted toward lower temperatures compared to $\beta_{\rm KWW}$.
Overall, consistent with the behavior of $\beta_{\rm KWW}$, finite-size effects appear at higher temperatures in smaller systems.

\section{Multiple relaxation modes picture of stretching exponent $\beta_{\rm KWW}$}\label{ap:multi}
In this appendix, we first review a phenomenological interpretation of stretched-exponential relaxation in terms of multiple relaxation modes, which we refer to as the multiple relaxation mode picture in this article.

The stretching exponent $\beta_{\rm KWW}$ is believed to take values smaller than unity when multiple relaxation modes with different timescales coexist~\cite{palmerModelsHierarchicallyConstrained1984}.
This picture can be expressed as
\begin{align}
\exp\left[-(t/\tau_\alpha)^{\beta_{\rm KWW}}\right]
= \int d\tau \exp(-t/\tau) P(\tau),
\label{eq:multi_mode_ap}
\end{align}
where $P(\tau)$ denotes the probability distribution of the relaxation time $\tau$.
Analytical derivations of Eq.~\ref{eq:multi_mode_ap} satisfying $0<\beta_{\rm KWW}<1$ have been demonstrated only for specific choices of $P(\tau)$, such as one-sided Lévy stable distributions~\cite{gorskaStretchedExponentialBehavior2017}.
Nevertheless, it is empirically well known that, when a distribution of relaxation times is present, relaxation functions can often be accurately fitted by a KWW-type stretched exponential form~\cite{yasudaDynamicRheologySupercooled2011}.
In the context of supercooled-liquid dynamics, analyses using the stretched exponential form have been performed in numerous studies~\cite{shangLocalGlobalStretched2019,berthierSelfInducedHeterogeneityDeeply2021,dasCrossoverDynamicsKobAndersen2022a}.
Assuming that Eq.~\ref{eq:multi_mode_ap} holds, one expects that the value of $\beta_{\rm KWW}$ is identical when the distributions $P(\tau/\langle\tau\rangle)$, normalized by the ensemble-averaged relaxation time $\langle\tau\rangle \equiv \int d\tau, \tau P(\tau)$, coincide.

\begin{figure*}[t]
\begin{center}
\includegraphics[width=.8\linewidth,angle=0]{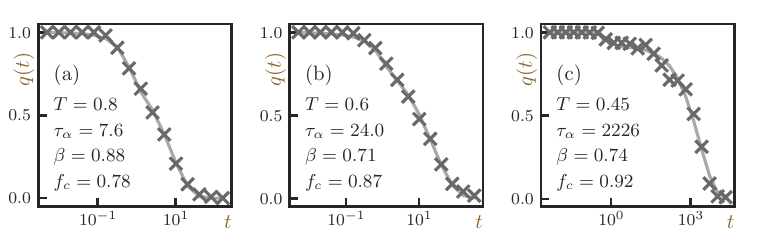}
\end{center}
\caption{
Semi-log plot of $q(t)$ as a function of the elapsed time $t$, obtained from a single relaxational dynamics trajectory in a system of size $N=1000$.
Results are shown for (a) $T=0.8$, (b) $T=0.6$, and (c) $T=0.45$.
Cross symbols denote the simulation results, and solid lines show the fitting curves based on the two-mode relaxation model, Eq.~\ref{eq:2mode}.
The fitted values of $\tau_\alpha$, $\beta_{\rm KWW}$, and $f_{\rm C}$ for each panel are summarized in the inset.}
\label{fig:single_relaxation}
\end{figure*}

\begin{figure*}[t]
\begin{center}
\includegraphics[width=.8\linewidth,angle=0]{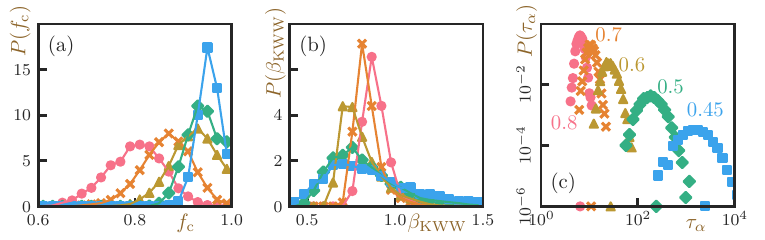}
\end{center}
\caption{
Probability distribution functions of the parameters of the two-mode relaxation model obtained from fitting $q(t)$ measured along single dynamical trajectories.
(a) slow-mode fraction $f_{\rm C}$,
(b) stretching exponent $\beta_{\rm KWW}$, and
(c) slow-mode relaxation time $\tau_\alpha$.
Each symbol corresponds to a different temperature, as indicated in panel (c).
The symbols shown on the horizontal axis of panel (c) denote the mean values of
$\tau_\alpha$ at the corresponding temperatures.
}
\label{fig:multi_modes}
\end{figure*}

\subsection{Relaxational dynamics of single trajectories}\label{ap:P_slow}
As discussed in Sec.~\ref{sec:Q}, the multiple relaxation modes picture provides a qualitative understanding of the dependence of the overlap function $Q(t)$ on temperature and system size obtained from molecular simulations.
However, this picture is based on phenomenological considerations and thus involves some subtlety.
To illustrate this subtlety in a concise manner, we present the overlap function $q(t)$ calculated along single trajectories.

For a system of size $N=1000$, we measured the time evolution of the overlap function $q(t)$ along single trajectories at each temperature (without averaging). 
Figure~\ref{fig:single_relaxation} shows representative single-trajectory relaxation behaviors at $T=0.8$, $0.6$, and $0.45$, together with the corresponding fits obtained using the two-mode model (Eq.~\ref{eq:2mode}).
These results indicate that even at the level of the overlap function $q(t)$ along a single trajectory, a two-step relaxation is already observed, and that, even when focusing solely on the slow mode, the relaxation follows a stretched-exponential functional form.
In other words, even when focusing exclusively on single trajectories, one cannot identify a single exponential relaxation mode that would provide a solid foundation for the multiple relaxation mode picture.
These results shown in Fig.~\ref{fig:single_relaxation} are expected to reflect the spatial heterogeneity of relaxation dynamics in supercooled liquids.
Indeed, Refs.~\cite{appignanesiReproducibilityDynamicalHeterogeneities2006,appignanesiDemocraticParticleMotion2006} have shown that structural relaxation in supercooled liquids proceeds via a sequence of localized, intermittent events.
Within the framework of Eq.~\ref{eq:multi_mode_ap}, such effects of spatial heterogeneity are not incorporated explicitly and only the temporal inhomogeneity is considered.
As a consequence, the present measurements do not allow for a direct verification of the discussion based on the multiple relaxation modes model presented in Sec.~\ref{sec:Q}.
Verification of the multiple relaxation modes picture using spatiotemporally resolved analysis remains an important subject for future work.

For reference, the probability distributions of $f_{\rm C}$, $\beta_{\rm KWW}$, and $\tau_\alpha$ computed from 3072 independent single trajectories are shown in Fig.~\ref{fig:multi_modes}(a-c).
We find that not only $\tau_\alpha$, but also $f_{\rm C}$ and $\beta_{\rm KWW}$ exhibit distributions at all temperatures investigated.

\section{Apparent activation energy}\label{ap:deltaE}

\begin{figure}[t]
\begin{center}
\includegraphics[width=\linewidth,angle=0]{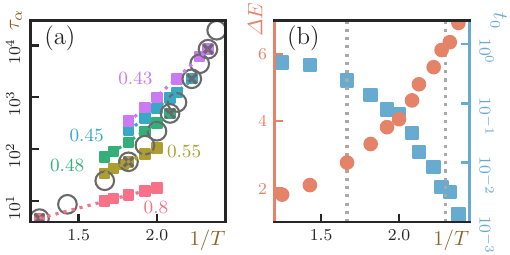}
\end{center}
\caption{
(a) Semi-log plot of the relaxation time as a function of the inverse temperature.
The nonequilibrium relaxation time $\tau_\alpha^{\rm RH}$ obtained by the reheating protocol and the corresponding Arrhenius fits are shown.
For representative values $T_{\rm p}=0.8$, $0.55$, $0.48$, $0.45$, and $0.43$, the reheating-protocol results $\tau_\alpha^{\rm RH}$ measured at several $T_{\rm s}$ are indicated by filled squares, while the fits to the Arrhenius law (Eq.~\ref{eq:Ar_RH}) are shown as dotted lines.
For reference, the equilibrium relaxation times $\tau_\alpha$ are indicated by open circles.
The crosses denote the equilibrium relaxation time $\tau_\alpha$ evaluated at the temperature at which the reheating-protocol data are presented.
(b) $\Delta E$ (circles, left axis) and $t_0$ (squares, right axis) obtained from the reheating protocol, plotted as functions of the inverse temperature. The two vertical dotted lines indicate the positions of $T_{\rm ava}\approx 0.6$ and $T_{\rm MCT}\approx 0.435$.
}
\label{fig:deltaE}
\end{figure}

In this section, we present the results for the apparent activation energy $\Delta E$ measured using the reheating protocol proposed in Ref.~\cite{picaciamarraLocalVsCooperative2024}.

\subsection{Constant approximation in ref.~\cite{coslovichDynamicThermodynamicCrossover2018a}}\label{ap:prev_deltaE}
The apparent activation energy in the KAM has already been measured by Coslovich \textit{et al}. in Ref.~\cite{coslovichDynamicThermodynamicCrossover2018a}.
However, in Ref.~\cite{coslovichDynamicThermodynamicCrossover2018a}, the activation energy was estimated using the following relation:
\begin{align}
\Delta E = \frac{d\ln \left(\tau_\alpha/t_0\right)}{d(1/T)},
\end{align}
where $t_0$ is a microscopic time scale.
This relation can be derived from the Arrhenius law (with the Boltzmann constant set to unity in our units),
\begin{align}
\tau_\alpha=t_0\exp\left(\frac{\Delta E}{T}\right)\Leftrightarrow \ln{\frac{\tau_\alpha}{t_0}}=\frac{\Delta E}{T}, \label{eq:deltaE_Arrhe}
\end{align}
by differentiating both sides with respect to $1/T$ under the assumption that all quantities except $\tau_\alpha$ are temperature independent.
In this procedure, the term $\frac{1}{T}\frac{d\Delta E}{d(1/T)}$, which arises when $\Delta E$ has an explicit temperature dependence, is neglected. 
Since the results reported in Ref.~\cite{coslovichDynamicThermodynamicCrossover2018a} indeed show that $\Delta E$ is an increasing function of $1/T$, this procedure, in principle, leads to an overestimation of $\Delta E$. 
This issue was pointed out in Ref.~\cite{picaciamarraLocalVsCooperative2024}.

\subsection{Reheating protocol}\label{ap:reheat}
Several methods for estimating the activation energy with improved accuracy have recently been proposed in Refs.~\cite{picaciamarraLocalVsCooperative2024,jiRoleExcitationsSupercooled2025}.
Here, we adopt the reheating protocol, which is the simplest among these methods.
In this method, equilibrium configurations prepared at the preparation temperature $T_{\rm p}$ are used as initial conditions for NVT simulations performed at different temperatures $T_{\rm s}$, which are hereafter referred to as the switching temperatures.
In this setting, the structural relaxation observed in the very early stage of aging, before time-translation invariance is established, is expected to be governed by the activation energy $\Delta E(T_{\rm p})$ associated with the preparation temperature.
Accordingly, as a quantitative measure of the relaxation process observed under the above protocol, we introduce a nonequilibrium overlap function as:
\begin{align}
Q^{\rm RH}(t;T_{\rm s}|T_{\rm p})=\langle q(t)\rangle^{\rm RH}_{T_{\rm s}|T_{\rm p}}.
\end{align}
Here, $\langle\cdot\rangle^{\rm RH}_{T_{\rm s}|T_{\rm p}}$ denotes a sample average taken over the initial relaxation processes observed at the switching temperature $T_{\rm s}$, starting from configurations prepared at the preparation temperature $T_{\rm p}$.
If the dynamics observed under the present reheating protocol is indeed governed by the activation energy $\Delta E(T_{\rm p})$, the relaxation time $\tau_\alpha^{\rm RH}(T_{\rm s}\mid T_{\rm p})$, defined by the condition $Q^{\rm RH}(\tau_\alpha^{\rm RH};T_{\rm s}\mid T_{\rm p})=1/e$ under different switching temperatures $T_{\rm s}$, is expected to obey the Arrhenius law
\begin{align}
\tau_\alpha^{\rm RH}(T_{\rm s}|T_{\rm p})=t_0(T_{\rm p})\exp(\Delta E(T_{\rm p})/T_{\rm s}).\label{eq:Ar_RH}
\end{align}
Here, the possible $T_{\rm p}$ dependence of $t_0$ is explicitly taken into account.
For swap systems, the values of $\Delta E(T_{\rm p})$ estimated using the reheating protocol have been shown to be quantitatively consistent with those obtained from other precise measurement methods proposed in Refs.~\cite{picaciamarraLocalVsCooperative2024,jiRoleExcitationsSupercooled2025}, and no $T_{\rm p}$ dependence of $t_0$ has been found.

The relaxation times $\tau_\alpha^{\rm RH}$ measured using the reheating protocol for the KAM with $N=1000$ are shown in Fig.~\ref{fig:deltaE}(a). 
Over the entire temperature range investigated, for each preparation temperature $T_{\rm p}$, the results obtained at different switching temperatures $T_{\rm s}$ are well fitted by the Arrhenius law, Eq.~\ref{eq:Ar_RH}.
The values of $\Delta E$ and $t_0$ obtained from the Arrhenius fitting are shown in Fig.~\ref{fig:deltaE}(b).
Consistent with Ref.~\cite{coslovichDynamicThermodynamicCrossover2018a}, the obtained values of $\Delta E$ increase upon cooling.
However, the values obtained at low temperatures are significantly smaller than those reported in Ref.~\cite{coslovichDynamicThermodynamicCrossover2018a}, where $8 \le \Delta E \le 16$ was reported in the range $0.5 \ge T \ge 0.4$.
As discussed in Sec.~\ref{ap:prev_deltaE}, this difference is consistent with expectations.

\subsection{Preparation temperature dependence of $t_0$}\label{ap:t0}
As shown in Fig.~\ref{fig:deltaE}(b), the reheating protocol also yields another indicator, the microscopic time scale $t_0$.
The time scale $t_0$ is found to decrease rapidly upon lowering the temperature.
By analogy with rate constants in chemical reactions, it is natural to expect a temperature dependence of $t_0$. 
In this framework, $t_0$ corresponds to a microscopic time scale set by the inverse of the frequency factor and is therefore expected to increase as the temperature decreases.
However, the results of $t_0$ in Fig.~\ref{fig:deltaE}(b) decrease upon cooling. 
This counterintuitive behavior may arise from estimating the activation energy based on macroscopic measurements that effectively incorporate multiple hopping processes characterized by different energy barriers. 
Moreover, given the major role of entropic effects discussed in Sec.~\ref{sec:t_avalanches}, a purely energetic treatment as in Eq.~\ref{eq:Ar_RH} may be insufficient to extract the underlying energy barrier height.
Therefore, it remains unclear at present to what extent the results of our measurements, including the values of $\Delta E$, are reliable.
We stress again that the results obtained using the reheating protocol for the swap system in Ref.~\cite{picaciamarraLocalVsCooperative2024} showed that $t_0$ remains constant and does not exhibit any temperature dependence.

To achieve more accurate measurements, it may be preferable to employ the SEER approach. 
SEER is a method that directly probes individual microscopic hopping events using temperature-varying inherent dynamics.
Using a similar approach based on inherent dynamics, Ref.~\cite{baity-jesiRevisitingConceptActivation2021} showed that the activation energy decreases upon cooling.
This trend is opposite to that reported in other studies of the activation energy~\cite{coslovichDynamicThermodynamicCrossover2018a,picaciamarraLocalVsCooperative2024,jiRoleExcitationsSupercooled2025} (note that refs.~\cite{picaciamarraLocalVsCooperative2024,jiRoleExcitationsSupercooled2025} studied the swap system).
As discussed in Secs.~\ref{sec:deloc} and \ref{sec:ubiquity}, even seemingly fundamental and universal properties, such as the localization of unstable modes and the non-Debye law, can exhibit qualitative differences across systems. 
The temperature dependence of the activation energy may likewise differ qualitatively across systems.
On the other hand, since Ref.~\cite{baity-jesiRevisitingConceptActivation2021} considered a very small system size ($N=65$), the possibility of finite-size effects cannot be excluded.
Thus, accurately characterizing the temperature dependence of the activation energy in the KAM remains an open issue.

\section{Influence of crystallization}\label{ap:crystal}
It has been reported that crystallization may become relevant at low temperatures in the KAM, even within the numerically accessible time window~\cite{ingebrigtsenCrystallizationInstabilityGlassForming2019a,coslovichDynamicThermodynamicCrossover2018a,dasCrossoverDynamicsKobAndersen2022a}.
In previous studies~\cite{coslovichDynamicThermodynamicCrossover2018a,dasCrossoverDynamicsKobAndersen2022a}, samples exhibiting significant crystallization were excluded from the analysis.
In the present study, we also assessed the possible influence of crystallization on the qualitative results.
To detect crystallized samples, we employed the common neighbor analysis (CNA) method~\cite{honeycuttMolecularDynamicsStudy1987,fakenSystematicAnalysisLocal1994}, which requires a relatively small number of free parameters, as also adopted in Ref.~\cite{coslovichDynamicThermodynamicCrossover2018a}.
Note that Ref.~\cite{dasCrossoverDynamicsKobAndersen2022a} employed a different crystallinity-detection scheme based on bond-level correlations of spherical harmonics.

\subsection{Crystallization detection}\label{ap:CNA}
In the KAM, the A particles, which constitute 80\% of the system, tend to form an fcc-type crystal structure at very low temperatures~\cite{pedersenPhaseDiagramKobAndersenType2018}.
Accordingly, following Ref.~\cite{coslovichDynamicThermodynamicCrossover2018a}, local crystallization was detected based on the fraction of CNA-142 bonds associated with the fcc structure.
A CNA-142 bond is defined as a bond between a pair of nearest-neighbor particles (as specified by the first digit, 1), for which the two bonded particles share four common neighbors (as specified by the second digit, 4), and among these four common neighbors there exist two bonds (as specified by the third digit, 2).
The degree of crystallinity can be quantified by the fraction of CNA-142 pairs, $f_{142}\equiv N_{142}/N_{\rm pairs}$, where $N_{142}$ denotes the number of CNA-142 pairs and $N_{\rm pairs}$ the total number of nearest-neighbor pairs in the system.

In this analysis, the only parameters to identify CNA-142 pairs are the neighbor cutoff distances $r_{\alpha\beta}^{\rm neighbor}$ ($\alpha,\beta\in{A,B}$).
In the present study, these cutoff distances were determined from the position of the first minimum of the radial distribution function $g(r)$ at the lowest temperature investigated, $T=0.41$, for the system size $N=1000$.
The resulting values are $r_{AA}^{\rm neighbor}=1.42$, $r_{AB}^{\rm neighbor}=1.27$, and $r_{BB}^{\rm neighbor}=1.10$ for the respective particle pairs.

\subsection{Detection of crystallized samples}\label{ap:detection}

We examine the parameter dependence of the fraction of CNA-142 bonds, $f_{142}$.
For each system size, the threshold for identifying crystallized samples, $f_{142}^{\rm C}(N)$, is defined as $f_{142}^{\rm C}(N)=1.2 f_{142}^\ast(N)$, where $f_{142}^\ast(N)$ denotes the maximum value of $f_{142}$ observed at $T=0.8$ for a given system size.
Samples satisfying $f_{142}>f_{142}^{\rm C}(N)$ are classified as crystallized.
The fraction of crystallized samples among all samples is denoted by $f_{\rm cry}$.

\begin{figure*}[t]
\begin{center}
\includegraphics[width=.8\linewidth,angle=0]{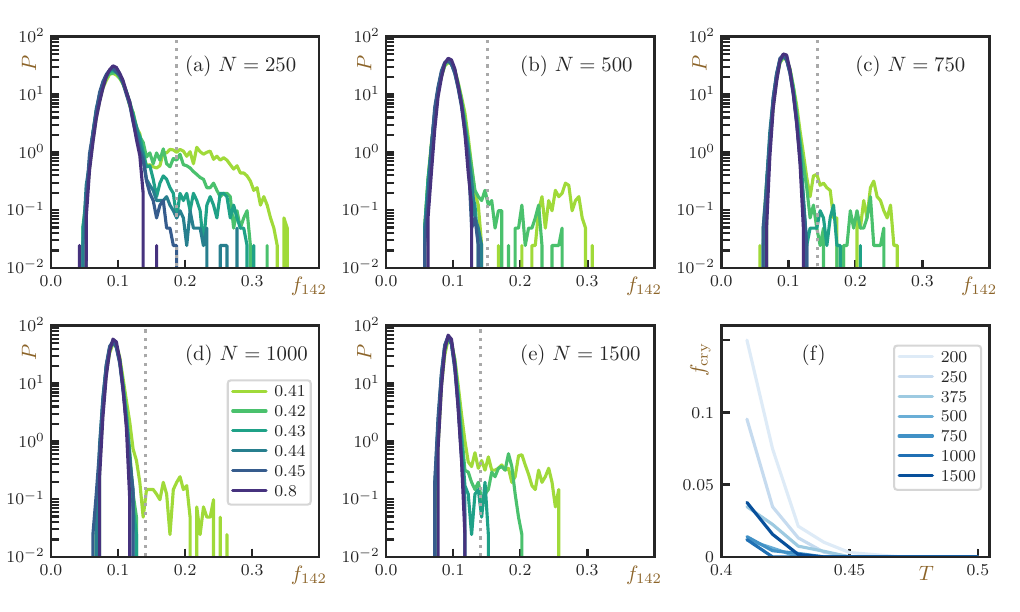}
\end{center}
\caption{
Probability distribution functions of the fraction of CNA-142 bonds, $f_{142}$.
(a)-(e) Semi-log plots of the results for system sizes $N=250$, $500$, $750$, $1000$, and $1500$.
In each panel, curves of different colors correspond to different temperatures, as indicated in the legend in panel (d).
Vertical dashed lines indicate the threshold values $f_{142}^{\rm C}$ used to identify crystallized samples for each system size.
(f) Linear plot of the fraction of crystallized samples, $f_{\rm cry}$, as a function of temperature. Curves of different colors correspond to different system sizes, as specified in the legend.}
\label{fig:CNA}
\end{figure*}

Figures~\ref{fig:CNA}(a)-(e) show the temperature dependence of the probability distribution functions of $f_{142}$ for different system sizes.
For all system sizes, $f_{142}$ is found to increase as the temperature is lowered.
In addition, we observe that the overall values of $f_{142}$ (and hence the threshold $f_{142}^{\rm C}$) tend to be larger for smaller system sizes.
Figure~\ref{fig:CNA}(f) presents the temperature and system-size dependence of the fraction of crystallized samples, $f_{\rm cry}$.
While $f_{\rm cry}$ generally decreases monotonically with increasing system size, the system with $N=1500$ exhibits a higher tendency to crystallize than those with $N=1000$ and $N=750$.

\begin{figure*}[t]
\begin{center}
\includegraphics[width=.8\linewidth,angle=0]{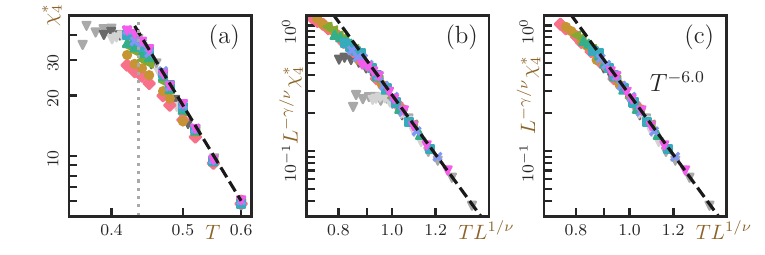}
\end{center}
\caption{
Comparison between data from previous studies~\cite{coslovichDynamicThermodynamicCrossover2018a,dasCrossoverDynamicsKobAndersen2022a} and our results for the peak value of the dynamical susceptibility, $\chi_4^\ast$.
Results obtained after removing crystallized samples are shown.
(a) Log-log plot of $\chi_4^\ast$ as a function of temperature.
(b) Finite-size scaling of $\chi_4^\ast$.
(c) Finite-size scaling of $\chi_4^\ast$ including only data points above the MCT transition temperature.
In all panels, only data for $T\le T_{\rm ava}$ are shown.
Dashed lines indicate the power-law behavior $\chi_4^\ast\sim T^{-\gamma}$, and the meanings of different symbols follow the legend in Fig.~\ref{fig:chi4_withprevious}.
For the data taken from Ref.~\cite{dasCrossoverDynamicsKobAndersen2022a}, the values are rescaled by a factor of $N/N_A=5/4$ to ensure consistency in the definition.
In panel (a), the vertical dashed line marks the position of $T_{\rm MCT}\approx 0.435$.}
\label{fig:CNA_test}
\end{figure*}

Figure~\ref{fig:CNA_test} shows the results of the analysis of the temperature and system-size dependence of $\chi_4^\ast$ after removing crystallized samples.
The meaning of each panel is the same as in Fig.~\ref{fig:chi4_withprevious}.
Even after excluding crystallized samples, within the temperature range investigated here, the qualitative conclusion remains unchanged: critical behavior emerges in a specific temperature window ($T_{\rm MCT}\le T\le T_{\rm ava}$).
Slight quantitative differences are observed in the low-temperature regime.

\section{Raw data of the vibrational density of states for stable samples}\label{ap:dos}
\begin{figure*}[t]
\begin{center}
\includegraphics[width=.8\linewidth,angle=0]{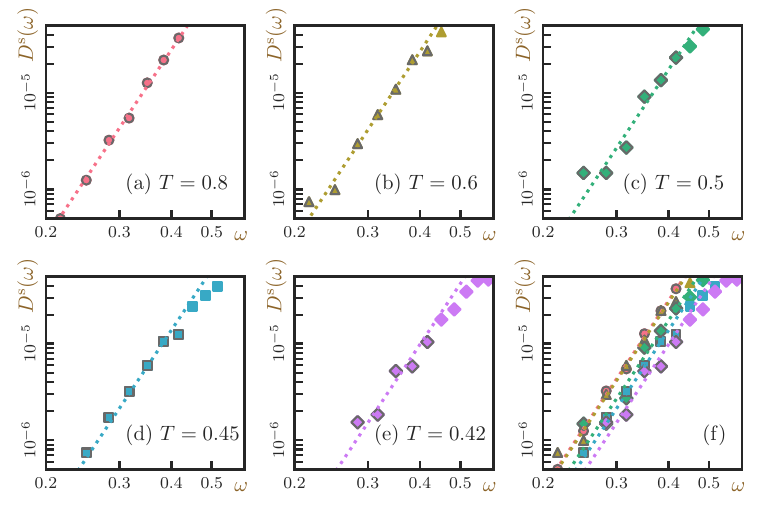}
\end{center}
\caption{
(a)-(e) Log-log plots of the low-frequency region of the vibrational density of states calculated from stable configurations only, $D^{\rm S}(\omega)$, shown as a function of the eigenfrequency $\omega$.
Each panel corresponds to a different temperature ($T=0.8$, $0.6$, $0.5$, $0.45$, and $0.42$), which are the same as those shown in Fig.~\ref{fig:vdos}(b) of the main text.
All results are for $N=1000$.
(f) Log-log plot of $D^{\rm S}(\omega)$ as a function of $\omega$, showing the results for all temperatures presented in panels (a)-(e) simultaneously.
Different symbols correspond to the temperatures indicated by the same symbols in panels (a)-(e).
In all panels, dotted lines indicate fits of the data to an $\omega^{6.5}$ power law at each temperature.}
\label{fig:raw_dos}
\end{figure*}

In Fig.~\ref{fig:vdos}(b) of the main text, the vibrational density of states calculated from stable configurations, $D^{\rm S}(\omega)$, was plotted as a function of a shifted frequency, $\omega_{\rm shift}$, where the shift along the eigenfrequency axis was introduced for better visual clarity.
In this Appendix, we present in Fig.~\ref{fig:raw_dos} the results for $D^{\rm S}(\omega)$ at each temperature as a function of the unshifted eigenfrequency $\omega$.

\bibliography{DH_INMs}

\end{document}